\documentclass[%
 reprint,
superscriptaddress,
 amsmath,amssymb,
 aps,
 nolongbibliography
]{revtex4-2}

\usepackage[final]{pdfpages}
\makeatletter
\AtBeginDocument{\let\LS@rot\@undefined}
\makeatother

\usepackage{graphicx}% Include figure files
\usepackage{dcolumn}% Align table columns on decimal point
\usepackage{bm}% bold math
\usepackage{hyperref}% add hypertext capabilities
\hypersetup{
	colorlinks   = true, %Colours links instead of ugly boxes
	urlcolor     = blue, %Colour for external hyperlinks
	linkcolor    = blue, %Colour of internal links
	citecolor    = blue %Colour of citations
}

\usepackage{array,multirow}
\usepackage{float}
\usepackage{subfigure}
\usepackage[export]{adjustbox}
\usepackage{xcolor}
\usepackage[T1]{fontenc}
\usepackage{verbatim}
\usepackage{booktabs}
\usepackage{natbib}

\DeclareMathAlphabet\mathbfcal{OMS}{cmsy}{b}{n}

\begin{document}

\preprint{APS/123-QED}

\title{Giant Enhancement of Vacuum Friction in Spinning YIG Nanospheres}

\author{Farhad Khosravi}
\affiliation{Department of Electrical and Computer Engineering, University of Alberta, Edmonton, Alberta T6G 1H9, Canada}
\affiliation{Elmore Family School of Electrical and Computer Engineering, Birck Nanotechnology Center, Purdue University, West Lafayette, Indiana 47907, USA}

\author{Wenbo Sun}
\affiliation{Elmore Family School of Electrical and Computer Engineering, Birck Nanotechnology Center, Purdue University, West Lafayette, Indiana 47907, USA}

\author{Chinmay Khandekar}
\affiliation{Elmore Family School of Electrical and Computer Engineering, Birck Nanotechnology Center, Purdue University, West Lafayette, Indiana 47907, USA}

\author{Tongcang Li}
\affiliation{Department of Physics and Astronomy, Purdue Quantum Science and Engineering Institute, Purdue University, West Lafayette, Indiana 47907, USA}
\affiliation{Elmore Family School of Electrical and Computer Engineering, Birck Nanotechnology Center, Purdue University, West Lafayette, Indiana 47907, USA}

\author{Zubin Jacob}
\email{zjacob@purdue.edu}
\affiliation{Elmore Family School of Electrical and Computer Engineering, Birck Nanotechnology Center, Purdue University, West Lafayette, Indiana 47907, USA}

\date{\today}
             
\begin{abstract}
Experimental observations of vacuum radiation and vacuum frictional torque are challenging due to their vanishingly small effects in practical systems. For example, a rotating nanosphere in free space slows down due to friction from vacuum fluctuations with a stopping time around the age of the universe. Here, we show that a spinning yttrium iron garnet (YIG) nanosphere near aluminum or YIG slabs exhibits vacuum radiation eight orders of magnitude larger than other metallic or dielectric spinning nanospheres. We achieve this giant enhancement by exploiting the large near-field magnetic local density of states in YIG systems, which occurs in the low-frequency GHz regime comparable to the rotation frequency. Furthermore, we propose a realistic experimental setup for observing the effects of this large vacuum radiation and frictional torque under experimentally accessible conditions.
\end{abstract}

\maketitle

\section{Introduction}
The physics of rotating nanoparticles is gaining more attention as recent technological advancements provide experimental platforms for rotating levitated nanoparticles at GHz speeds~\cite{kani2022magnonic,reimann2018ghz,ahn2018optically,piotrowski2023simultaneous,gonzalez2021levitodynamics,jin20216,PhysRevA.108.063511,doi:10.1021/acs.nanolett.3c02442}. Besides having implications in the fields of quantum gravity~\cite{marletto2017gravitationally}, dark energy detection~\cite{rider2016search}, and superradiance~\cite{bekenstein1998superradiance}, rotating nanoparticles are crucial for studying the effects of quantum vacuum fluctuations~\cite{deop2023control,guo2021single,pan2019circular,xu2020enhancement,hummer2020acoustic,pan2019magnetically}. Rotating nanoparticles can emit real photons and experience frictional torques from the fluctuating quantum vacuum even at zero temperature~\cite{manjavacas2010vacuum,maghrebi2012spontaneous}. Although Casimir forces between static objects have been measured extensively~\cite{decca2003measurement,mohideen1998precision,munday2009measured}, the experimental sensitivity is only starting to reach the limit needed to measure the frictional torque exerted on rotating nanoparticles from the vacuum~\cite{xu2017detecting}. Meanwhile, direct observation of vacuum radiation from rotating nanoparticles remains challenging due to the extremely low number of radiated photons.

In the specific case of moving media or rotating particles, a unique regime of light-matter interaction occurs when the material resonance frequency becomes comparable to the mechanical motion frequency~\cite{pendharker2017pt,guo2014giant,guo2014singular}. In particular, a giant enhancement or even a singularity is possible in vacuum fluctuation effects~\cite{pendharker2017pt,guo2014giant,guo2014singular}. Recently, world record rotation frequencies were achieved for optically levitated nanospheres~\cite{reimann2018ghz,ahn2018optically,jin20216}. This immediately opens the question of whether unique material resonances comparable to this rotation frequency can help enter a new regime of light-matter interaction. Here, we show that gyromagnetic yttrium iron garnet (YIG) exhibits the magnon polariton resonance at GHz frequencies~\cite{hisatomi2016bidirectional,zhang2014strongly} comparable to the levitated nanoparticle's rotation frequency, providing a unique opportunity for enhancing vacuum fluctuation effects on rotating nanoparticles.

In this article, we put forth an approach to enhance and observe the vacuum radiation and frictional torques by leveraging a YIG nanosphere spinning at $\Omega=1\,\mathrm{GHz}$ in the vicinity of a metallic or YIG interface. Our proposal exploits an asymmetry between the electric and magnetic local density of states (LDOS) which was previously reported in Ref.~\cite{joulain2003definition}. In particular, near conventional metals, the electric LDOS is enhanced at optical frequencies, whereas the magnetic LDOS becomes dominant at GHz frequencies. Therefore, our proposal exploits magnetic materials with magnon polaritons to enhance the magnetic local density of states beyond those of conventional plasmonic metals. Due to the large magnetic LDOS and YIG magnetic resonance at GHz frequencies, the fluctuating magnetic dipoles of the YIG nanosphere can strongly couple to a large density of evanescent waves in the near-field of metallic and magnetic interfaces, leading to colossal vacuum radiation. 

We demonstrate that a spinning YIG nanosphere generates vacuum radiation eight orders of magnitude larger than other metallic or dielectric nanospheres in the vicinity of a metallic or magnetic slab. We show that, near magnetic materials, most of this radiated energy can be transferred to surface magnon polaritons. Furthermore, we reveal that the large vacuum radiation and vacuum friction have experimentally observable effects on the balance rotation speed, stopping time, and balance temperature of the spinning YIG nanospheres under experimentally accessible rotation speeds, particle sizes, temperatures, and vacuum pressures. Therefore, the setup proposed in this article based on spinning YIG nanospheres represents a unique tool for detecting and analyzing vacuum radiation and frictional torques.

\section{Giant Vacuum Radiation from Spinning YIG Nanospheres}

\begin{figure}[!tp]
    \centering
    \includegraphics[width =3.4 in]{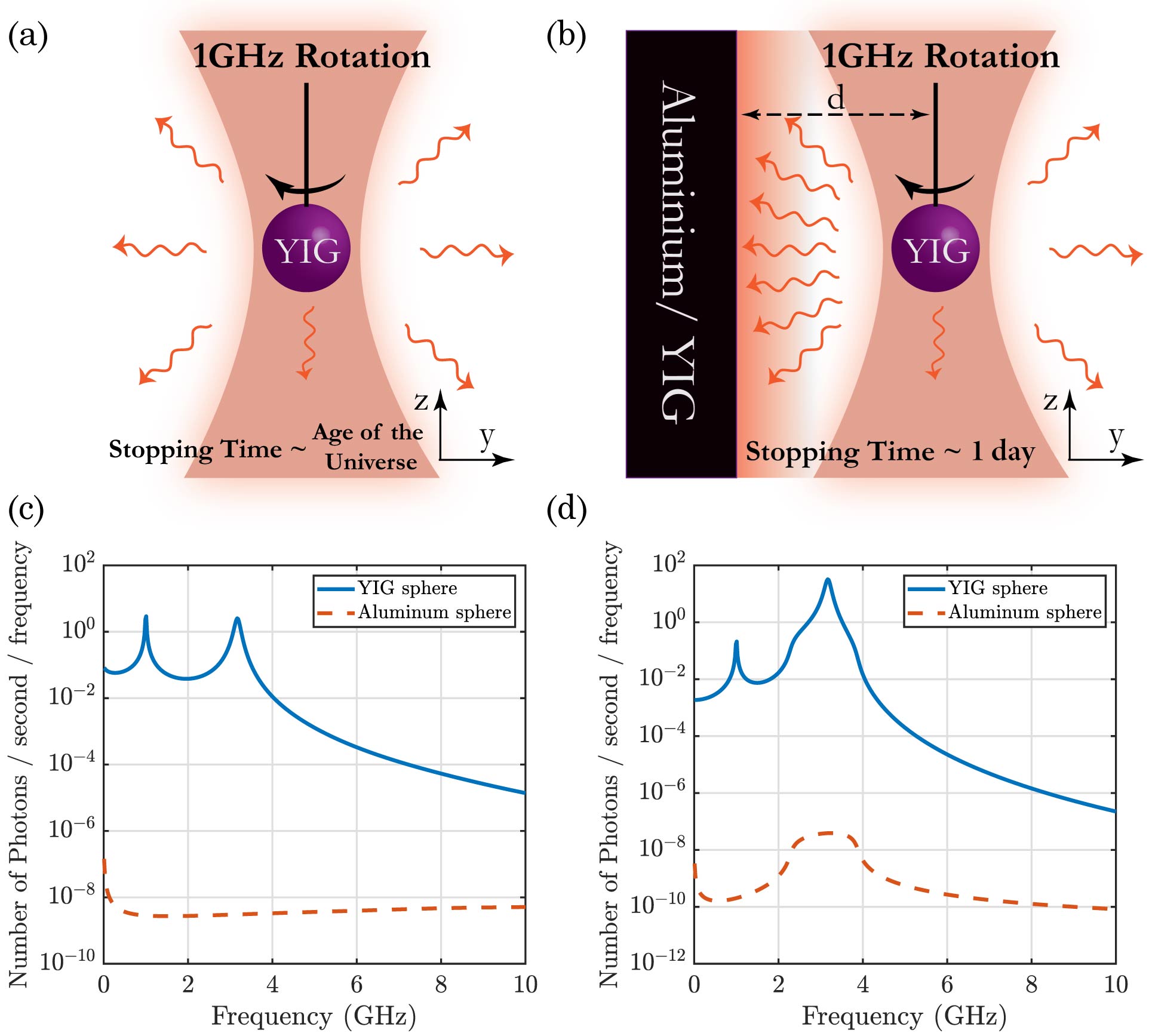}
    \caption{(a) A YIG sphere trapped in the laser beam and spinning at 1 GHz rotation frequency in the vacuum. The stopping time for the sphere is on the order of the age of the universe. (b) YIG sphere spinning in the vicinity of an Aluminum or YIG interface exhibits colossal vacuum radiation. The stopping time, due to the presence of the interface, is reduced to about 1 day. (c, d) Number of photons emitted per second per radiation frequency, defined as $\frac{1}{\hbar \omega} dP/d\omega = \Gamma(\omega) - \Gamma(-\omega)$, for a YIG (blue solid curve) or Aluminum (dashed orange curve) nanosphere of radius 200 nm at distance $d=0.5\,\mathrm{\mu m}$ from (c) an aluminum slab or (d) a YIG slab at room temperatures. For the Al slab, a non-local model has been used. The YIG slab in panel (d) is biased along the y direction (panel (a)) with a magnetic field of $H_0=812$ Oe.}
    \label{Fig:Schematic_&_Radiation}
\end{figure}

We first consider the vacuum radiation from a spinning YIG nanosphere with a radius of $200\,\mathrm{nm}$, as illustrated in Fig.~\ref{Fig:Schematic_&_Radiation}(a,~b). A stationary nanosphere at the equilibrium temperature exhibits zero net radiation since the number of photons emitted by the fluctuating dipoles of the nanosphere is equal to the number of photons absorbed by the nanosphere from the fluctuating electromagnetic fields in the vacuum. However, for rotating nanospheres, the balance between the emitted and absorbed photons is broken. A net radiated power from the nanosphere arises even at zero temperature due to the extra boost of mechanical rotational energy~\cite{manjavacas2010thermal}. The source of this vacuum radiation energy is the non-inertial motion of the nanosphere, which is transferred to generate real photons from vacuum fluctuations~\cite{maghrebi2012spontaneous}. Based on fluctuational electrodynamics (see derivations in Appendix~\ref{Sec:Vacuum_Radiation}), we find the total radiated power from a spinning YIG nanosphere $P^{\text{rad}} = \int_0^\infty d\omega \hbar \omega \left[ \Gamma^\text{H}(\omega) - \Gamma^\text{H}(-\omega) \right]$ can be determined from $\Gamma^\text{H}(\omega)$, which is the spectral density of the radiation power arising from magnetic dipole fluctuations.

In the absence of any interface, vacuum radiation from a spinning YIG nanosphere does not exhibit any substantial enhancement. However, metallic or magnetic interfaces can drastically change this observation. Metallic nanospheres are known to possess higher radiation rates compared to dielectric nanospheres near material interfaces~\cite{zhao2012rotational,manjavacas2010thermal}. Here, we observe that magnetic nanospheres exhibit even larger radiation rates, which are about eight orders of magnitude compared to metallic nanospheres near metallic or magnetic interfaces, as shown in Fig.~\ref{Fig:Schematic_&_Radiation}(c,~d). We demonstrate that radiated photons per second per frequency expressed through $\Gamma^\text{H}(\omega) - \Gamma^\text{H}(-\omega)$ from spinning YIG nanospheres (blue curves) are much more than those from the aluminum nanospheres (orange curves) near Al interfaces (Fig.~\ref{Fig:Schematic_&_Radiation}(c)) and YIG interfaces (Fig.~\ref{Fig:Schematic_&_Radiation}(d)). Furthermore, we find that a spinning YIG nanosphere radiates about $6$ femtowatts of power, in stark contrast to the Al sphere, which radiates about $6\times 10^{-7}$ femtowatts near Al interfaces (Fig.~\ref{Fig:Schematic_&_Radiation}(c)). In the vicinity of YIG interfaces (Fig.~\ref{Fig:Schematic_&_Radiation}(d)), we find about $61.3$ femtowatts and $4.63\times 10^{-7}$ femtowatts of radiated power from YIG and Al nanospheres, respectively. The radiated energy mostly goes into the lossy surface waves in both metallic and magnetic materials~\cite{FORD1984195}. However, if the magnetic material is properly biased, as is the case studied here with a bias magnetic field of $812$ Oe for the YIG slab, the magnetic resonance in the magnetic slab can become resonant with the magnetic resonance in the magnetic sphere. As a result, most of the radiated energy is transferred to surface magnon polaritons. These results clearly show the advantage of YIG over Al nanospheres for probing vacuum radiation.

The above results are explained by the YIG magnon polariton resonance at GHz frequencies and differences in the low-frequency electric and magnetic LDOS near metallic and magnetic interfaces. Vacuum fluctuation effects on rotating nanoparticles can be significantly enhanced when the rotation frequency is comparable to resonance frequencies. In addition, as shown by Joulain \textit{et al.}~\cite{joulain2003definition}, LDOS near metals is dominated by the magnetic LDOS at wavelengths above a few microns. Here, we extend this observation to magnetic materials and take into account the effects of non-local electromagnetic response in Al~\cite{FORD1984195} (also see Appendix~\ref{Sec:nonlocalal}). Higher magnetic LDOS than electric LDOS at low frequencies originates from differences in the reflection of the $s$- and $p$-polarized evanescent waves. The near-field electric LDOS is mainly influenced by $p$-polarized evanescent waves since their contributions to the electric LDOS are strongly momentum-dependent and dominate the high momentum contributions crucial for near-field LDOS. In contrast, the opposite is true for the near-field magnetic LDOS, and the contributions from the $s$-polarized evanescent waves dominate. At GHz frequencies, the imaginary part of the reflection coefficient for evanescent $s$-polarized waves is much larger than that for evanescent $p$-polarized waves. Thus, the $s$ polarization contributes more to the LDOS than the $p$ polarization, leading to a more dominant magnetic LDOS near metallic and magnetic interfaces. These near-field LDOS can be further enhanced by material resonances~\cite{pendharker2017pt,guo2014giant,guo2014singular,volokitin2007near,sloan2021casimir}. 

To this end, we discuss the spectral density $\Gamma^\text{H}(\omega)$ that determines the vacuum radiation. Through a similar approach as the methods used by Abajo and Manjavacas~\cite{manjavacas2010vacuum}, our result for the radiation spectral density $\Gamma^\text{H}(\omega)$ of a spinning gyromagnetic nanosphere due to magnetic dipole fluctuations is (see derivations in Appendix~\ref{Sec:Vacuum_Radiation}):
\begin{widetext}
\begin{equation}\label{Eq:Spetral_density_xz}
  \begin{split}
  \Gamma^\text{H}(\omega) = (\omega \rho_0/8) \Bigg\{ & \left[g_{\bot,2}^\text{H}(\omega) + 2g_{\parallel}^\text{H}(\omega) + 2g_{g,2}^\text{H}(\omega) \right] \left[ \text{Im}\left\{\alpha_{m,\bot}(\omega^-)\right\} - \text{Re}\left\{ \alpha_{m,g}(\omega^-) \right\} \right] \left[n_1(\omega^-) - n_0(\omega)\right] \\
  + & g_{\bot,1}^\text{H}(\omega) \text{Im}\left\{ \alpha_{m,\parallel}(\omega) \right\}\left[n_1(\omega) - n_0(\omega)\right] \Bigg\},
  \end{split}
\end{equation}
\end{widetext}
where $\rho_0 = \omega^2/c^2\pi^3$ is the vacuum density of states, $g_{\bot,1}^\text{H}$ , $g_{\bot,2}^\text{H}$ are the two components of the magnetic Green's function in the plane of the interface (the $xx$ and $zz$ components for the setup shown in Fig.~\ref{Fig:Schematic_&_Radiation}(b)), $g_\parallel^\text{H}$ is the component normal to the interface (the $yy$ component here), and $g_{g,2}^\text{H}$ is the off-diagonal component between the in-plane and normal directions (the $xy$ component here), all normalized by $\pi\omega \rho_0/8$. $\alpha_{m,\bot}(\omega)$, $\alpha_{m,g}(\omega)$, and $\alpha_{m,\parallel}(\omega)$ are the $xx$ (or $yy$), $xy$, and $zz$ components of the YIG nanosphere magnetic polarizability tensor in the rotating sphere frame (see Appendix~\ref{Sec:Polarizability_tensor} for derivations). $\Omega$ is rotating frequency of the nanosphere and $\omega^- = \omega - \Omega$. $n_1(\omega)$ and $n_0(\omega)$ are the Bose-Einstein distribution functions pertinent to the sphere temperature $T_1$ and the environment temperature $T_0$, respectively. Detailed derivations for all these quantities and discussions of various YIG interface orientations and bias magnetic field directions are provided in Appendix~\ref{Sec:Green_function}. When the sphere is stationary $\omega^- = \omega$, and the sphere temperature is equal to the temperature of the environment $T_1 = T_0$, the terms $n_1(\omega^-) - n_0(\omega)$ and $n_1(\omega) - n_0(\omega)$ become zero; thus, the radiation becomes zero as expected.

Here, we emphasize one important aspect of $\Gamma^\text{H}(\omega)$ regarding the rotation-induced magnetization of the YIG nanosphere, which can occur without any external magnetic field. This is known as the Barnett effect and originates from the conservation of angular momentum, where the mechanical angular momentum of the sphere is transferred to the spin of the unpaired electrons in the magnetic material~\cite{barnett1915magnetization}. Assuming the magnetic field is parallel to the rotation axis, the Larmor precession frequency $\omega_0$ of the electrons inside the sphere is~\cite{pozar2006microwave} (also see Appendix~\ref{Sec:barnett}):
\begin{equation}\label{Eq:Larmor Freuqnecy}
    \omega_0 = \Omega + \mu_0 \gamma H_0,
\end{equation}
for the electron gyromagnetic ratio $\gamma$, vacuum permeability $\mu_0$, and applied external magnetic field $H_0$. We incorporate this effect on $\omega_0$ to find the magnetic response of the spinning YIG nanosphere.

\section{Enhancement of Vacuum frictional torque}
We now discuss the vacuum frictional torque exerted on the rotating YIG nanosphere in the vicinity of YIG and Al interfaces. We use a similar approach to find the vacuum torque exerted on the spinning gyromagnetic YIG sphere due to magnetic dipole and magnetic field fluctuations (detailed derivations are provided in Appendix~\ref{Sec:torque}). The torque along the axis of rotation is given by $M_z = \int_0^\infty d\omega \hbar \left[\Gamma_\text{M}^\text{H}(\omega) + \Gamma_\text{M}^\text{H}(-\omega)\right]$, where the expression for $\Gamma_\text{M}^\text{H}(\omega)$ is similar to the expression for $\Gamma^\text{H}(\omega)$ in Eq.~(\ref{Eq:Spetral_density_xz}), with the difference being that the last term on the second line is not present in $\Gamma^\text{H}_\text{M}(\omega)$ (see Appendix~\ref{Sec:torque}). Additionally, we find that other components of the torque ($M_x$ and $M_y$ components) are not necessarily zero in the vicinity of the YIG interface, in contrast to the Al slab. Due to the anisotropy of the YIG slab, $M_x$ and $M_y$ do not vanish for some directions of the bias magnetic field. We provide further discussions of these cases in the supplementary material.

\begin{figure}[t!]
    \centering
    \includegraphics[width=3.4 in]{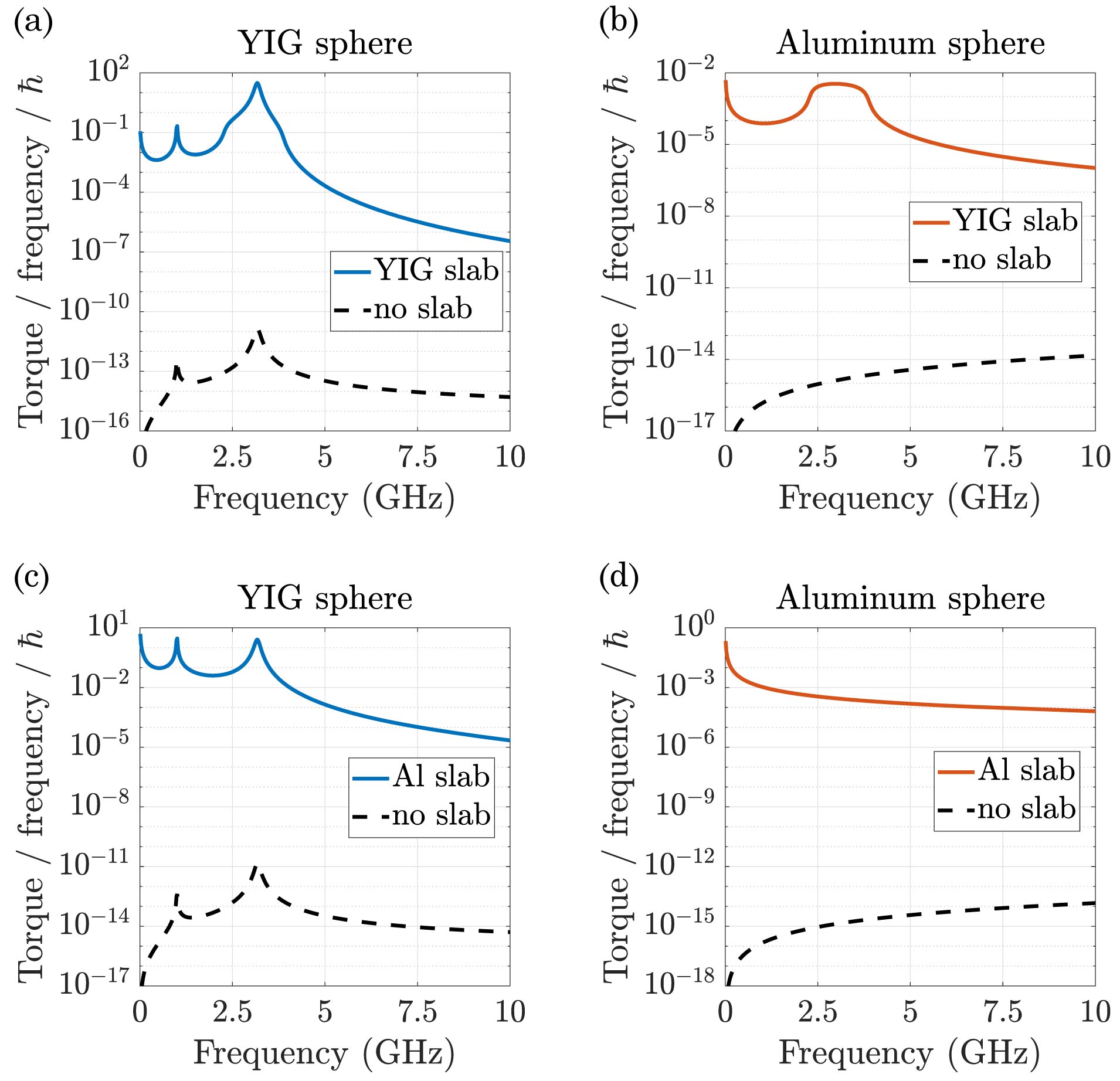}
    \caption{The negative vacuum frictional torque experienced by the YIG and aluminum nanosphere with a radius of 200 nm at room temperature. (a) Torque experienced by a YIG sphere in the vicinity of the YIG slab (solid blue curve) and in vacuum (dashed black curve). (b) Torque exerted on an Al sphere in the vicinity of the YIG slab (solid orange curve) and in vacuum (dashed black curve). (c), (d) the same as (a) and (b) with the YIG slab replaced by an Al slab. The YIG slab is biased along the $y$ direction with $H_0=812$ Oe (see Fig. 1(a)). A non-local model is used for the Al slabs. The distance between the spinning spheres and slabs is $d=0.5 \mu$m for all cases. Placing the YIG or Al interface in the vicinity of spinning nanospheres results in about 12 orders of magnitude increase in the exerted vacuum torque.}
    \label{Fig:Torques}
\end{figure}

In Fig.~\ref{Fig:Torques}, we compare vacuum torques exerted on spinning YIG nanospheres (Fig.~\ref{Fig:Torques}(a,~c)) and spinning Al nanospheres (Fig.~\ref{Fig:Torques}(b,~d)), on nanospheres spinning in the vicinity of YIG slabs (Fig.~\ref{Fig:Torques}(a,~b)) and Al slabs (Fig.~\ref{Fig:Torques}(c,~d)), as well as on nanospheres spinning in the vicinity of slabs (solid colored curves) and spinning in vacuum (dashed black curves). We demonstrate that vacuum torques exhibit more than 10 orders of magnitude enhancement in the vicinity of YIG and Al slabs compared to the vacuum, and about 4 orders of magnitude enhancement due to employing YIG nanospheres instead of Al nanospheres. These results unravel the advantage of utilizing YIG nanospheres for probing vacuum frictional torques at GHz frequencies. In Fig.~\ref{Fig:Torques}, we consider non-local electromagnetic response~\cite{FORD1984195} for Al interfaces and incorporate effects from the magnetic and electric dipole and field fluctuations on vacuum torques. We notice that the vacuum torque is dominated by magnetic rather than electric fluctuations in all cases (see Appendix~\ref{Sec:torque}). In addition, we have taken into account the effect of recoil torque~\cite{nieto2015optical} --  the torque exerted on the sphere due to the scattering of vacuum field fluctuations off the particle. As discussed in Appendix~\ref{Sec:torque}, we find that effects from this second-order torque are negligible compared with the effects of magnetic fluctuations in the studied cases.

\section{Observable Outcomes of Giant Vacuum Friction in Spinning YIG nanospheres}
\begin{figure}[t!]
    \centering
    \includegraphics[width=3.4 in]{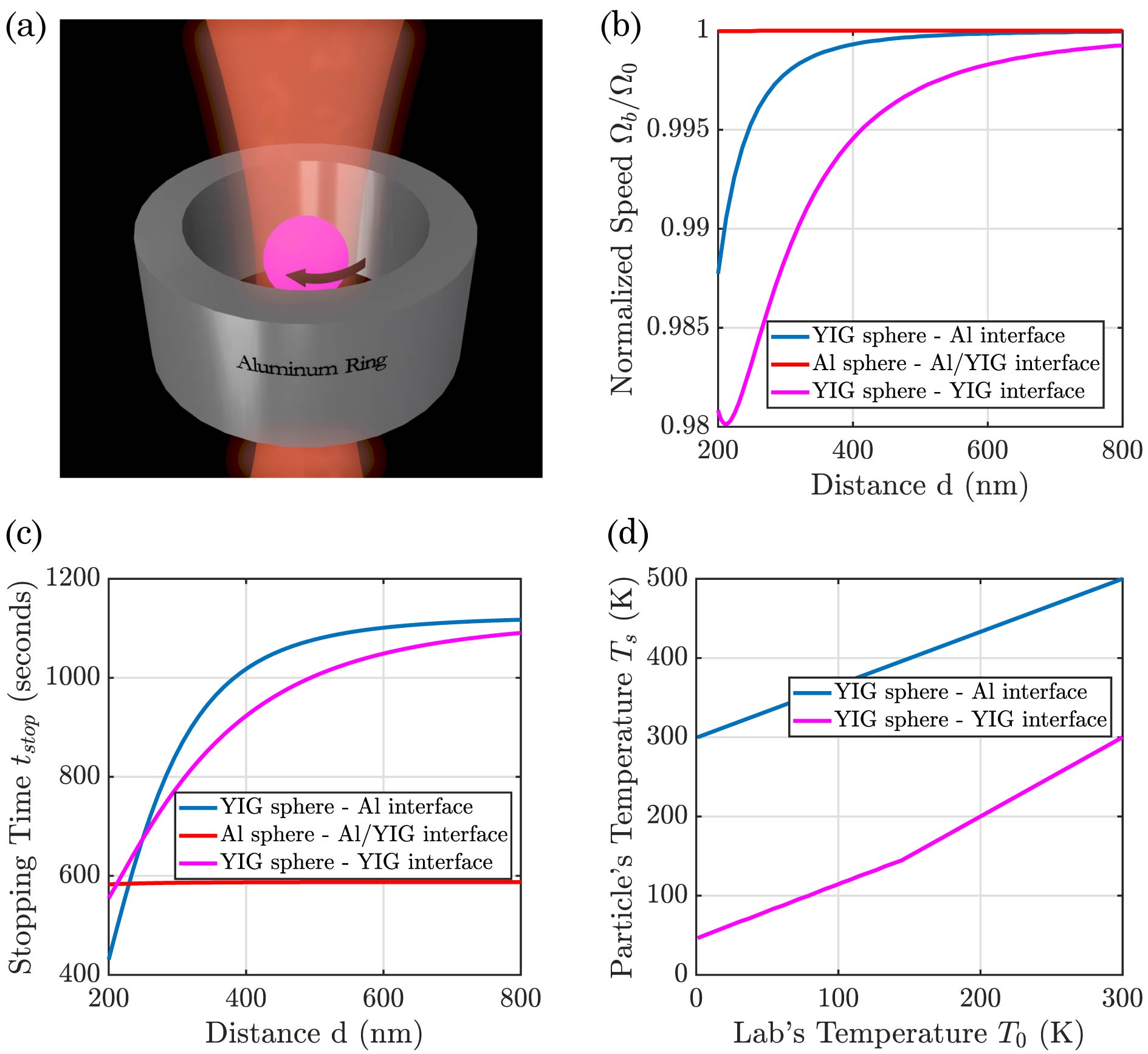}
    \caption{Experimental considerations of the setup. (a) Proposed experimental setup with nanosphere trapped inside a ring. (b) Balance rotation speed $\Omega_b$ for Al sphere (red curve) and YIG sphere in the presence of Al (blue curve) and YIG (pink curve) interfaces, as a function of distance $d$ from the interface for a $200$ nm  radius sphere at $10^{-4}$ Torr vacuum pressure. The values are normalized by the vacuum balance rotation speed $\Omega_0$. (c) Characteristic stopping time as a function of distance from the interface at $10^{-6}$ Torr vacuum pressure. (d) Balance temperature of the YIG sphere $T_s$ at $d=500$ nm distance from Al (blue curve) and YIG (pink curve) interfaces as a function of lab temperature $T_0$, at $10^{-4}$ Torr vacuum pressure. For Al spheres, there is no final temperature as the temperature keeps rising with time.} 
    \label{Fig:Experimental_Considerations}
\end{figure}

The observable effects of the colossal vacuum radiation and frictional torques come down to changes in experimentally measurable parameters when the spinning nanosphere is brought closer to the vicinity of Al/YIG interfaces. In Fig.~\ref{Fig:Experimental_Considerations}(a), we show the proposed experimental setup for this observation where a YIG nanosphere is trapped inside an Al or YIG ring. We note that the size of the ring is much larger than that of the nanosphere, and it does not lead to any resonant behavior. However, for smaller ring sizes, LDOS can be further enhanced compared to the slab interface case due to the presence of interfaces on all sides. 

We evaluated some observable experimental outcomes due to large vacuum radiation and friction. This analysis is based on the experimentally accessible parameters from Refs. \cite{ahn2018optically,seberson2019optical,seberson2019distribution}. In Fig.~\ref{Fig:Experimental_Considerations}(b), we show the balanced rotation speed $\Omega_b$ of the spinning nanosphere normalized by the rotation speed $\Omega_0$ in the absence of any interface as a function of distance $d$ from the interface. The balance rotation speed is defined as the sphere's stable, perpetual rotation speed and occurs when the driving force due to the laser is equal to the drag force due to the vacuum chamber. In the absence of any interface, due to the negligible value of vacuum radiation, the balance rotation speed $\Omega_0$ is obtained when the torque from the trapping laser balances the frictional torque from air molecules in the imperfect vacuum~\cite{ahn2018optically} (also see Apeendix~\ref{Sec:exptana}). We assume the laser driving torque is constant and the drag force from air molecules has a linear dependence on rotational speeds~\cite{ahn2018optically}. In Fig.~\ref{Fig:Experimental_Considerations}(b), we show that the balance rotation speed of the YIG nanosphere is reduced when it is closer to Al (blue curve) or YIG (pink curve) interfaces, as a result of the large frictional torques from vacuum fluctuations. Remarkably, we notice that there is no observable change in the balance speed for spinning Al nanospheres in the vicinity of Al or YIG interfaces (red curve). 

In Fig.~\ref{Fig:Experimental_Considerations}(c,~d), we further demonstrate outcomes of the large vacuum radiation in other experimental observables, such as the stopping time as a function of distance (Fig.~\ref{Fig:Experimental_Considerations}(c)) and the balance temperature as a function of the vacuum temperature $T_0$ (Fig.~\ref{Fig:Experimental_Considerations}(d)). Stopping time is the time constant of the exponential decrease of the nanosphere rotation velocity after the driving torque is turned off. The torque can be switched off by changing the polarization of the trapping laser from circular to linear without having to switch off the trapping laser. The balance temperature refers to the nanosphere temperature $T_s$, at which the loss of mechanical rotational energy due to vacuum frictional torque stops heating the nanospheres. As shown in Fig.~\ref{Fig:Experimental_Considerations}(c,~d), YIG nanospheres exhibit distinct behaviors in the stopping time and balance temperature compared to Al nanospheres near YIG and Al interfaces.

The results of Fig.~\ref{Fig:Experimental_Considerations} show that the vacuum radiation and frictional torque can be experimentally measured through the balance speed, balance temperature, and stopping time of the YIG nanosphere. In stark contrast, the Al nanosphere (or any other metallic nanospheres) may not experience enough vacuum friction to exhibit observable outcomes unless it is in a sensitive setup with very low vacuum pressure~\cite{ahn2018optically,xu2017detecting}. 

\section{Discussion and Conclusion}
Our results show that due to YIG magnon polariton resonance and the dominance of magnetic LDOS over electric LDOS in the vicinity of metallic or magnetic materials at GHz frequencies, spinning YIG nanospheres can exhibit orders of magnitude larger vacuum radiation and frictional torque compared to any metallic or dielectric nanosphere. By investigating the case of a YIG nanosphere spinning at 1 GHz speed, we have shown that the effect of colossal vacuum fluctuations can be observed in an experimentally accessible setup. Our results set a new perspective for observing and understanding radiation and frictional torques from vacuum fluctuations. Furthermore, our discussions of magnetic LDOS near YIG interfaces under various bias fields pave the way for magnetometry~\cite{sloan2019controlling} and spin measurement~\cite{Mamin2013nanoscale} applications.

\section*{Acknowledgements}
This research was supported by the Army Research Office under grant number W911NF-21-1-0287 and the Office of Naval Research under award number N000142312707.

\appendix

\section{Radiation Power due to Magnetic Fluctuations}\label{Sec:Vacuum_Radiation}
In this appendix, we provide detailed derivations of the radiation power $P^\mathrm{rad}$ from a spinning YIG nanosphere and its spectral density $\Gamma^\text{H}(\omega)$ due to magnetic fluctuations. Using an approach similar to that taken by Abajo \textit{et. al}~\cite{manjavacas2010thermal,manjavacas2010vacuum}, we can write the radiated power due to the magnetic fluctuations of dipoles and fields as,
\begin{equation}\label{Eq:Radiated_power_Magnetic_01}
  P_{\text{mag}} = - \langle \bm H^{\text{ind}}\cdot \partial \bm m^{\text{fl}}/\partial t + \bm H^{\text{fl}}\cdot \partial \bm m^{\text{ind}}/\partial t \rangle,
\end{equation}
where $\bm H^\text{ind}$ is the induced magnetic field due to the magnetic dipole fluctuations $\bm m^\text{fl}$ of the particle and $\bm m^\text{ind}$ is the induced magnetic dipole in the particle due to the fluctuations of the vacuum magnetic field $\bm H^\text{fl}$. Note that all of these quantities are written in the lab frame. For the sphere spinning at the rotation frequency $\Omega$, we can write,
\begin{equation}
    \begin{split}
        m_x^\text{fl} = & m_x^{'\text{fl}} \cos \Omega t - m_y^{'\text{fl}} \sin \Omega t, \\
        m_y^\text{fl} = & m_x^{'\text{fl}} \sin \Omega t + m_y^{'\text{fl}} \cos \Omega t, \\
        m_z^\text{fl} = & m_z^{'\text{fl}},
    \end{split}
\end{equation}
where the primed quantities are written in the rotating frame. Performing a Fourier transform as $\bm m^{'\text{fl}}(t) = \int \frac{d\omega}{2\pi}e^{-i\omega t} \bm m^{'\text{fl}}(\omega)$, we can write in the frequency domain
\begin{equation}\label{Eq:m_lab_to_rotating}
    \begin{split}
        m_x^\text{fl}(\omega) = & \frac{1}{2}\left[m_x^{'\text{fl}}(\omega^-) + m_x^{'\text{fl}}(\omega^+) + i m_y^{'\text{fl}}(\omega^+) - im_y^{'\text{fl}}(\omega^-)\right], \\
        m_y^\text{fl}(\omega) = & \frac{1}{2}\left[i m_x^{'\text{fl}}(\omega^-) - i m_x^{'\text{fl}}(\omega^+) + m_y^{'\text{fl}}(\omega^+) + m_y^{'\text{fl}}(\omega^-)\right],
    \end{split}
\end{equation}
where $\omega^\pm = \omega \pm \Omega$. We can similarly write for the magnetic fields
\begin{equation}\label{Eq:H_rotating_to_lab}
    \begin{split}
        H_x^{'\text{fl}}(\omega) = & \frac{1}{2}\left[H_x^\text{fl}(\omega^+) + H_x^\text{fl}(\omega^-) - i H_y^\text{fl}(\omega^+) + i H_y^\text{fl}(\omega^-)\right],\\
        H_y^{'\text{fl}}(\omega) = & \frac{1}{2}\left[i H_x^\text{fl}(\omega^+) - i H_x^\text{fl}(\omega^-) + H_y^\text{fl}(\omega^+) + H_y^\text{fl}(\omega^-) \right].
    \end{split}
\end{equation}
Thus, using the fact that,
\begin{equation}\label{Eq:m_vs_Hfl}
    \bm m^{'\text{ind}}(\omega) = \Bar{\bm \alpha}_m(\omega) \cdot \bm H^{'\text{fl}}(\omega),
\end{equation}
with
\begin{equation}
    \bar{\bm \alpha}_m(\omega) = \begin{pmatrix} 
    \alpha_{m,\bot}(\omega) & -\alpha_{m,g}(\omega) & 0 \\
    \alpha_{m,g}(\omega) & \alpha_{m,\bot}(\omega) & 0 \\
    0 & 0 & \alpha_{m,\parallel}(\omega) 
    \end{pmatrix}
\end{equation}
being the magnetic polarizability tensor of the YIG sphere biased along the $z$ axis, we find in the lab frame
\begin{equation}
    \bm m^\text{ind}(\omega) = \bar{\bm \alpha}_m^\text{eff}(\omega) \cdot \bm H^\text{fl}(\omega),
\end{equation}
\newpage
\onecolumngrid
where
\begin{subequations}
\begin{equation}
    \bar{\bm \alpha}_m^\text{eff} = \begin{pmatrix}
    \alpha_{m,\bot}^\text{eff}(\omega) & -\alpha_{m,g}^\text{eff}(\omega) & 0 \\
    \alpha_{m,g}^\text{eff}(\omega) & \alpha_{m,\bot}^\text{eff}(\omega) & 0 \\
    0 & 0 & \alpha_{m,\parallel}^\text{eff}(\omega) 
    \end{pmatrix},
\end{equation}
  \begin{equation}\label{Eq:alpha_bot_eff}
    \alpha_{m,\bot}^\text{eff}(\omega) = \frac{1}{2}\left[ \alpha_{m,\bot}(\omega^+) + \alpha_{m,\bot}(\omega^-) + i\alpha_{m,g}(\omega^+) - i\alpha_{m,g}(\omega^-) \right],
  \end{equation}
  \begin{equation}
    \alpha_{m,g}^\text{eff}(\omega) = -\frac{i}{2}\left[ \alpha_{m,\bot}(\omega^+) - \alpha_{m,\bot}(\omega^-) + i\alpha_{m,g}(\omega^+) + i\alpha_{m,g}(\omega^-) \right],
  \end{equation}
  \begin{equation}\label{Eq:alpha_parallel_eff}
    \alpha_{m,\parallel}^\text{eff} = \alpha_{m,\parallel}(\omega).
  \end{equation}
\end{subequations}
Note that we have used an expression similar to Eq.~(\ref{Eq:m_lab_to_rotating}) but written for the induced magnetic dipole moments. Expression for $\alpha_{m,\bot}(\omega)$ and $\alpha_{m,g}(\omega)$ are given in Appendix \ref{Sec:Polarizability_tensor}. Using the fluctuation-dissipation theorem (FDT)~\cite{landau2013statistical},
\begin{equation}
 \langle H^{\text{fl}}_i(\omega) H^{\text{fl}}_j(\omega')
 \rangle = 4\pi\hbar [n_0(\omega)+ 1]\left\{\frac{ G^\text{H}_{ij}(\omega) - G_{ji}^{\text{H}*}(\omega) }{2i}\right\}\delta(\omega + \omega'),
\end{equation}
with $G_{ij}^\text{H}(\omega) = G_{ij}^\text{H}(\bm r,\bm r'=\bm r,\omega)$ defined as the equal-frequency magnetic Green's function of the environment defined through the equation,
\begin{equation}
    H_i(\bm r,\bm r',\omega) = G_{ij}^\text{H}(\bm r,\bm r',\omega) m_j(\bm r',\omega),
\end{equation}
we find the second term in Eq.~(\ref{Eq:Radiated_power_Magnetic_01}) employing Eqs.~(\ref{Eq:H_rotating_to_lab}) and (\ref{Eq:m_vs_Hfl}):
\begin{equation}\label{Eq:Radiation_Magnetic_FlcFields_01}
  \begin{split}
  \langle H_i^\text{fl} (\omega) \partial m_i^\text{ind}(\omega')/\partial t \rangle = -i\omega' & 2\pi \hbar \Big[ n_0(\omega) + 1 \Big]\delta(\omega + \omega') \times \\
  \Bigg\{ & \left( \text{Im}\left\{ G_{xx}^\text{H}(\omega) \right\} + \text{Im}\left\{ G_{yy}^\text{H}(\omega) \right\} \right)\left[ \alpha_{m,\bot}(\omega^{'+}) + \alpha_{m,\bot}(\omega^{'-}) + i\alpha_{m,g}(\omega^{'+}) - i\alpha_{m,g}(\omega^{'-}) \right] \\
  + & \left(  \text{Re}\left\{G_{xy}^\text{H}(\omega)\right\} - \text{Re}\left\{ G_{yx}^{*\text{H}}(\omega) \right\} \right) \left[ \alpha_{m,\bot}(\omega^{'+}) - \alpha_{m,\bot}(\omega^{'-}) + i\alpha_{m,g}(\omega^{'+}) + i\alpha_{m,g}(\omega^{'-}) \right] \\
  + & 2 \text{Im}\left\{ G_{zz}^\text{H}(\omega) \right\} \alpha_{m,\parallel}(\omega') \Bigg\},
\end{split}
\end{equation}
where $n_0(\omega) = 1/(e^{\hbar \omega /k_\text{B}T_0}-1)$ is the Planck distribution at the temperature of the lab $T_0$. Writing FDT for the fluctuating dipoles, 
\begin{equation}
  \langle m_i^{'\text{fl}} (\omega)m_j^{'\text{fl}}(\omega') \rangle = 4\pi \hbar [n_1(\omega) +1 ]\left(\frac{\alpha_{m,ij}(\omega) - \alpha_{m,ji}^*(\omega) }{2i}\right) \delta(\omega +\omega'),
\end{equation}
we find the first term in Eq.~(\ref{Eq:Radiated_power_Magnetic_01}) employing Eq.~(\ref{Eq:m_lab_to_rotating}) and $H_i^\text{ind}(\omega) = G_{ij}^\text{H}(\omega) m_j^\text{fl}(\omega)$:
\begin{equation}\label{Eq:Radiation_Magnetic_FlcDipoles_01}
  \begin{split}
     \langle H_i^\text{ind}(\omega) \partial m_i^\text{fl}(\omega')/\partial t \rangle = &  - 2\pi\hbar i\omega'\left[n_1(\omega^-) + 1\right] \Bigg\{ \delta(\omega + \omega') \Big[ G_{xx}^\text{H}(\omega) \text{Im}\left\{ \alpha_{m,\bot}(\omega^-) \right\} - G_{xx}^\text{H}(\omega) \text{Re}\left\{ \alpha_{m,g}(\omega^-) \right\} \\
     - & G_{yy}^\text{H}(\omega) \text{Re}\left\{ \alpha_{m,g}(\omega^-) \right\} + G_{yy}^\text{H}(\omega) \text{Im}\left\{ \alpha_{m,\bot}(\omega^-) \right\} + iG_{xy}^\text{H}(\omega) \text{Im}\left\{ \alpha_{m,\bot}(\omega^-) \right\} \\
     - & iG_{xy}^\text{H}(\omega) \text{Re}\left\{ \alpha_{m,g}(\omega^-) \right\} + iG_{yx}^\text{H}(\omega) \text{Re}\left\{ \alpha_{m,g}(\omega^-) \right\} - iG_{yx}^\text{H}(\omega) \text{Im}\left\{ \alpha_{m,\bot}(\omega^-) \right\} \Big] \Bigg\} \\
     - & 2\pi\hbar i\omega' \left[n_1(\omega^+) + 1\right] \Bigg\{   \delta(\omega + \omega') \Big[G_{xx}^\text{H}(\omega) \text{Im}\left\{ \alpha_{m,\bot}(\omega^+)\right\} + G_{xx}^\text{H}(\omega) \text{Re}\left\{ \alpha_{m,g}(\omega^+) \right\}\\
     + & G_{yy}^\text{H}(\omega) \text{Re}\left\{ \alpha_{m,g}(\omega^+) \right\} + G_{yy}^\text{H}(\omega) \text{Im}\left\{ \alpha_{m,\bot}(\omega^+) \right\} - iG_{xy}^\text{H}(\omega) \text{Im}\left\{ \alpha_{m,\bot}(\omega^+) \right\} \\
     - & iG_{xy}^\text{H}(\omega) \text{Re}\left\{ \alpha_{m,g}(\omega^+) \right\} + iG_{yx}^\text{H}(\omega) \text{Re}\left\{ \alpha_{m,g}(\omega^+) \right\} + iG_{yx}^\text{H}(\omega) \text{Im}\left\{ \alpha_{m,\bot}(\omega^+) \right\} \Big] \Bigg\} \\
     - & 4\pi\hbar i\omega'\left[n_1(\omega) + 1 \right] \Bigg\{\delta(\omega + \omega') G_{zz}^\text{H}(\omega) \text{Im}\left\{ \alpha_{m,\parallel}(\omega) \right\} \Bigg\},
  \end{split}
\end{equation}
where $n_1(\omega)$ is the Planck distribution at the sphere temperature $T_1$. Taking the inverse Fourier transform, adding Eqs.~(\ref{Eq:Radiation_Magnetic_FlcFields_01}) and (\ref{Eq:Radiation_Magnetic_FlcDipoles_01}), taking the real part of the radiated power, and changing integral variables, we find
\begin{equation}\label{Eq:P_mag_anisotropic_xy}
  \begin{split}
    P_\text{mag} = \frac{\hbar}{\pi} \int_{-\infty}^{+\infty} \omega d\omega \Bigg\{ &\left[n_1(\omega^-)- n_0(\omega)\right] \left( \text{Im}\left\{ G_{xx}^\text{H}(\omega) \right\} + \text{Im}\left\{ G_{yy}^\text{H}(\omega) \right\} + \text{Re}\left\{ G_{xy}^\text{H}(\omega) \right\} - \text{Re}\left\{ G_{yx}^\text{H}(\omega) \right\} \right) \times \\
    & \left[ \text{Im}\left\{\alpha_{m,\bot}(\omega^{-})\right\} - \text{Re}\left\{\alpha_{m,g}(\omega^{-})\right\} \right] + \left[n_1(\omega) - n_0(\omega)\right] \text{Im}\left\{G_{zz}^\text{H}(\omega)\right\} \text{Im}\left\{\alpha_{m,\parallel}^\text{H}(\omega)\right\} \Bigg\}.
  \end{split}
\end{equation}
In this derivation, we have used the property $\alpha_{m}(-\omega) = \alpha_m^*(\omega)$. The expressions for Green's functions in different YIG and aluminum interface arrangements are given in Appendix~\ref{Sec:Green_function}. Plugging these expressions into Eq.~(\ref{Eq:P_mag_anisotropic_xy}), we obtain Eq.~(\ref{Eq:Spetral_density_xz}) in the main text. 

\section{Green's Function Near an Anisotropic Magnetic Material}\label{Sec:Green_function}

\begin{figure}[t!]
    \centering
    \includegraphics[width=3.4 in]{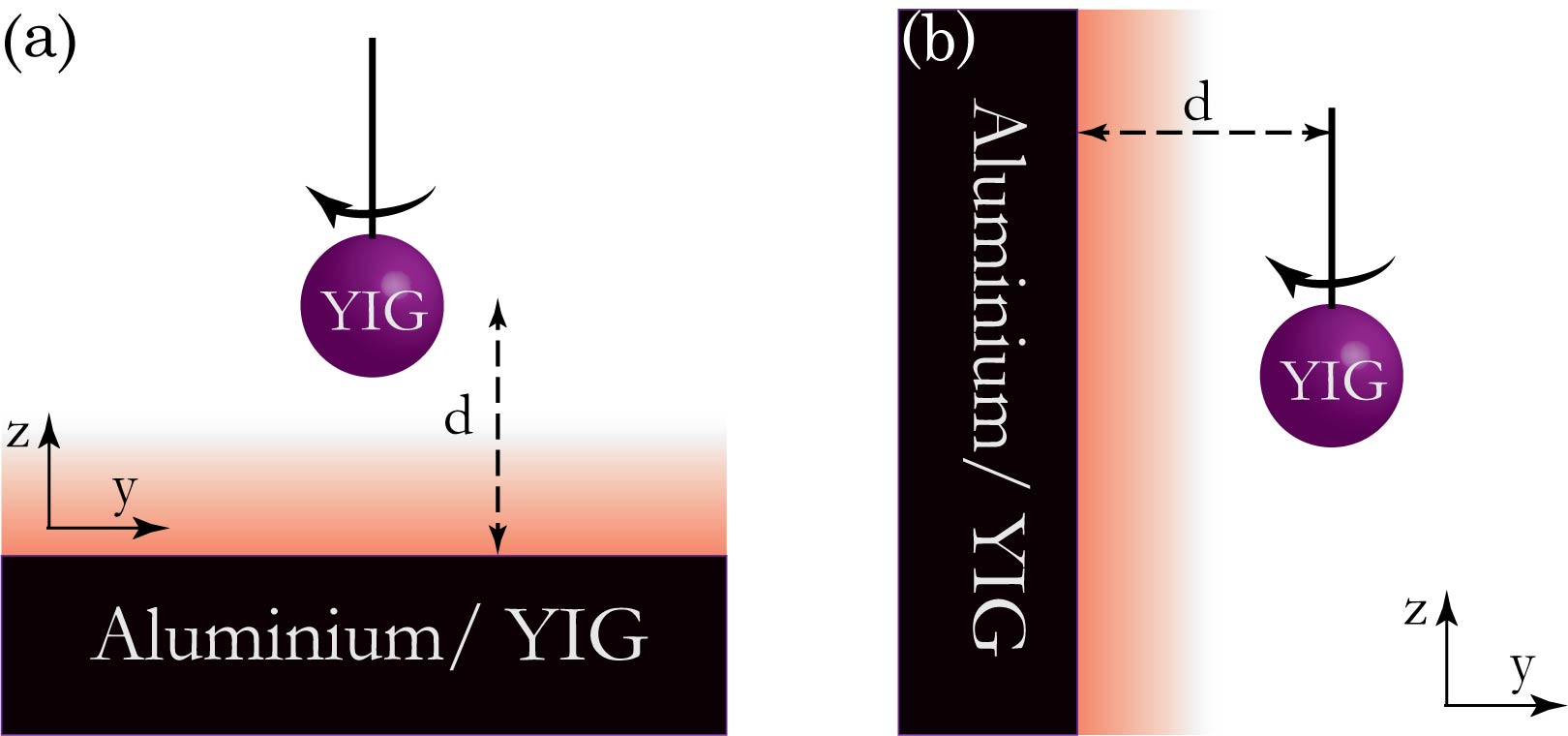}
    \caption{Schematic of the problem for the two cases of when the interface is in (a) $x-y$ plane and (b) $x-z$ plane.}
    \label{Fig:Schematic_XY_XZ}
\end{figure}

In this appendix, we provide the Green's function near a half-space of magnetic material, which would change due to the anisotropy of the material. We study two cases when the interface is the $x-y$ plane and $x-z$ plane, as shown in Fig.~\ref{Fig:Schematic_XY_XZ}. We can write the electric and magnetic fields in the vacuum as
\begin{subequations}
\begin{equation}
  \bm E = \bm E^i + \bm E^r, \quad \bm H = \bm H^i  + \bm H^r,
\end{equation}
\begin{equation}
  \bm E^i = (E_{0s} \hat{s}_- + E_{0p}\hat{p}_-)e^{i\bm k_-\cdot \bm r},
\end{equation}
\begin{equation}
  \bm E^r = \left( E_{0s} r_{ss} \hat{s}_+ + E_{0p} r_{pp} \hat{p}_+ + E_{0s} r_{ps} \hat{p}_+ + E_{0p} r_{sp} \hat{s}_+ \right)e^{i\bm k_+ \cdot \bm r},
\end{equation}
\begin{equation}
  \bm H^i = \frac{1}{\eta_0}\left(  - E_{0s} \hat{p}_- + E_{0p} \hat{s}_- \right)e^{i \bm k_- \cdot \bm r},
\end{equation}
\begin{equation}
  \bm H^r = \frac{1}{\eta_0}\left( - E_{0s} r_{ss} \hat{p}_+ + E_{0p} r_{pp} \hat{s}_+ + E_{0s}r_{ps} \hat{s}_+ - E_{0p} r_{sp} \hat{p}_+ \right) e^{i \bm k_+ \cdot \bm r},
\end{equation}
\end{subequations}
where $\hat{s}_\pm$, $\hat{p}_\pm$, and $\hat{k}_\pm/k_0$ form a triplet with
\begin{equation}\label{B2}
  \hat{k}_\pm = k_0 \left( \kappa \cos\phi \hat{x} + \kappa \sin\phi \hat{y} \pm k_z \hat{z} \right), \quad \hat{s}_\pm = \sin \phi \hat{x} - \cos \phi \hat{y}, \quad \hat{p}_{\pm} = -\left( \pm k_z \cos\phi \hat{x} \pm k_z \sin \phi - \kappa \hat{z} \right),
\end{equation}
and $\eta_0 = \sqrt{\mu_0/\epsilon_0}$, $k_0=\omega/c$, $\kappa^2 + k_z^2=1$, and $k_0k_z$ is the $z$ component of the wavevector. Similarly, we can write the electric and magnetic fields inside the magnetic material as
\begin{subequations}
  \begin{equation}
    \bm E' = \bm E^t, \quad \bm H' = \bm H^t,
  \end{equation}
  \begin{equation}
    \bm E^t = \left( E_{0s}t_{ss} \hat{s}'_- + E_{0p} t_{pp} \hat{p}'_- + E_{0s} t_{ps} \hat{p}'_- + E_{0p}t_{sp} \hat{s}'_- \right)e^{i\bm k'_-\cdot r},
  \end{equation}
  \begin{equation}
    \bar{\bar{\bm \mu}} \bm H^t = \frac{\sqrt{\kappa^2 + k_z'^2}}{\eta_0} \left[ - E_{0s} t_{ss} \hat{p}'_- + E_{0p} t_{pp} \hat{s}'_- +  E_{0s}t_{ps} \hat{s}'_- - E_{0p} t_{sp} \hat{p}'_- \right]e^{i\bm k'_- \cdot \bm r },
  \end{equation}
\end{subequations}
where
\begin{equation}\label{Eq:basis_vectors}
  \bm{k}'_\pm = k_0  \hat{k}'_\pm = k_0 \left(\kappa \cos\phi \hat{x} + \kappa \sin\phi \hat{y} \pm k_z' \hat{z}\right), \quad \hat{s}'_\pm = \sin \phi \hat{x} - \cos \phi \hat{y}, \quad \hat{p}'_{\pm} = - \frac{ \pm k_z' \cos\phi \hat{x} \pm k_z' \sin \phi \hat{y} - \kappa \hat{z}}{\sqrt{\kappa^2 + k_z'^2}}.
\end{equation}
Note that $\kappa$ is the same in the two media due to the boundary conditions. Also $\hat{k}'_\pm \times \hat{p}'_\pm = \hat{s}'_\pm$. We can write Maxwell's equations in the magnetic material in matrix form as~\cite{khandekar2019thermal}
\begin{equation}\label{Eq:Eq_of_Motion_inYIG}
  \left( M + M_k \right)\psi = \left[\begin{pmatrix}
                                 \bar{\bar{\bm \epsilon}} & 0 \\
                                 0 & \bar{\bar{\bm \mu}}
                               \end{pmatrix} + \begin{pmatrix}
                                 0 & \bar{\bar{\bm \kappa}} \\
                                 -\bar{\bar{\bm \kappa}} & 0
                               \end{pmatrix}
                               \right]\begin{pmatrix}
                                 \bm E^t \\
                                 \eta_0 \bm H^t
                               \end{pmatrix} = 0,
\end{equation}
where
\begin{equation}
  \bar{\bar{\bm \kappa}} = \begin{pmatrix}
                             0 & -k_z' & \kappa\sin\phi \\
                             k_z' & 0 & - \kappa \cos\phi \\
                             -\kappa \sin\phi & \kappa \cos\phi & 0
                           \end{pmatrix}.
\end{equation}
Setting the $\text{det}(M + M_k) = 0$ we get the solutions for $k_z'$ in terms of $\kappa$ and $\phi$~\cite{khandekar2019thermal}. From these solutions and applying the boundary conditions, we can find the values of $r_{ss}$, $r_{sp}$, $r_{ps}$, $r_{pp}$ for a given $\kappa$ and $\phi$. Note that different bias directions for the magnetic field of the YIG slab change the $\bar{\bar{\mu}}$ tensor and thus change the reflection coefficients $r_{ss}$, $r_{sp}$, $r_{ps}$, $r_{pp}$. %This can be done since no assumption on the direction and components of the permeability tensor has been assumed in our derivations. 

% \begin{figure}[t!]
%      \centering
%      \subfloat[\label{Fig:Schematic_XY}]{
%     \includegraphics[width=0.18\textwidth]{Schematic_xy.eps}}
%     \subfloat[\label{Fig:Schematic_XZ}]{
%     \includegraphics[width=0.18\textwidth]{Schematic_xz.eps}}
%     \caption{Schematic of the problem for the two cases of when the interface is in (a) $x-y$ plane and (b) $x-z$ plane.}
%     \label{Fig:Schematic_XY_XZ}
% \end{figure}

In the following, we first provide the expression for the magnetic dyadic Green's function $\bar{\bm G}^\text{H}$ for a source at $z' = d$ when the interface is in the $x-y$ plane (Fig.~\ref{Fig:Schematic_XY_XZ}(a)). Here, we take the spinning sphere to be at the origin to simplify the derivations and move $z = 0$ to $z' = d$. This would not change the Fresnel reflection coefficients. The incident magnetic Green's function at the location of the source is thus,
\begin{equation}
  \bar{\bm G}^\text{H}_i(z = z', \omega) = \frac{ik_0^2}{8\pi^2 \epsilon_m} \int \frac{dk_x dk_y}{k_z}(\hat{s}\hat{s} + \hat{p}_- \hat{p}_-) e^{ik_x(x - x') + ik_y(y -y')}.
\end{equation}
The reflected magnetic Green's function at the location of the source is
\begin{equation}
  \bar{\bm G}^\text{H}_r(z=z',\omega) = \frac{ik_0^2}{8\pi \epsilon_m} \int \frac{dk_x dk_y}{k_z}\left( \hat{s}r_{pp} \hat{s} + \hat{p}_+ r_{sp}\hat{s} + \hat{s}r_{ps}\hat{p}_- + \hat{p}_+ r_{ss} \hat{p}_- \right)e^{2ik_z d},
\end{equation}
where $k_x = \kappa \cos\phi$ and $k_y = \kappa \sin \phi$. Note that here the Fresnel reflection coefficients generally depend on the incidence angle $\phi$. For the special case of magnetization along the $z$ axis, they become independent of $\phi$. Using Eq.~(\ref{Eq:basis_vectors}) and dropping the terms that vanish after integration over $\phi$, we can write the total magnetic Green's function at the location of source as,
\begin{equation}\label{Eq:Greens_function_anisotropic}
    \begin{split}
       \bar{\bm G}^\text{H} (\bm r,\bm r, \omega) =  & \frac{ik_0^3}{8\pi^2} \int_{0}^{2\pi} d\phi \int_{0}^{+\infty} \frac{\kappa d\kappa}{p} \Bigg\{ \left[\sin^2\phi \hat{x}\hat{x} + \cos^2\phi \hat{y}\hat{y} - \sin\phi\cos\phi \left(\hat{x}\hat{y} + \hat{y}\hat{x}\right)\right]\left( 1 + r_{pp}e^{2ik_z d} \right) \\
      + & p^2\cos^2\phi \hat{x}\hat{x} + p^2\sin^2\phi \hat{y}\hat{y} + \kappa^2\hat{z}\hat{z} \\
      + & e^{2i k_0 p d}r_{ss} \big[- p^2 \cos^2\phi \hat{x}\hat{x} - p^2 \sin^2\phi\hat{y}\hat{y} + \kappa^2\hat{z}\hat{z} - p^2\cos\phi \sin\phi \left( \hat{x}\hat{y} + \hat{y}\hat{x} \right) \\
      - & p\kappa \cos\phi \left(\hat{x}\hat{z} - \hat{z}\hat{x}\right) - p\kappa \sin\phi \left(\hat{y}\hat{z} - \hat{z}\hat{y}\right)\big] \\
      + & e^{2ik_0 pd}r_{ps}\left[ p\sin\phi \cos\phi  \left(\hat{x}\hat{x} - \hat{y}\hat{y} \right) + p \sin^2\phi \hat{x}\hat{y} - p\cos^2\phi \hat{y}\hat{x} + \kappa \sin\phi \hat{x}\hat{z} - \kappa \cos\phi \hat{y}\hat{z} \right] \\
      + & e^{2ik_0 pd} r_{sp} \left[ -p\cos\phi \sin\phi \left(\hat{x}\hat{x} - \hat{y}\hat{y}\right) + p\cos^2\phi \hat{x}\hat{y} - p\sin^2\phi \hat{y}\hat{x} + \kappa\sin\phi \hat{z}\hat{x} - \kappa \cos\phi \hat{z}\hat{y} \right] \Bigg\}.
  \end{split}
\end{equation}
Note that the electric Green's function can be obtained by changing $r_{ss}$ to $r_{pp}$, $r_{pp}$ to $r_{ss}$, $r_{ps}$ to $r_{sp}$ and $r_{sp}$ to $r_{ps}$ and dividing by $\epsilon_0$. In general, the non-diagonal parts of the Green's function are not zero. Using this equation, we find,
\begin{subequations}\label{Eq:G_tensor_components}
\begin{equation}\label{Eq:G_xx_XYplane}
  \text{Im}\left\{G_{xx}^\text{H}(\omega)\right\} = \frac{\pi \omega \rho_0}{8} g_{\bot,1}^\text{H}(\omega),
\end{equation}
\begin{equation}\label{Eq:G_yy_XYplane}
  \text{Im}\left\{G_{yy}^\text{H}(\omega)\right\} = \frac{\pi \omega \rho_0}{8} g_{\bot,2}^\text{H}(\omega),
\end{equation}
\begin{equation}\label{Eq:G_xy_re_XYplane}
  \text{Re}\left\{G_{xy}^\text{H}(\omega)\right\} - \text{Re}\left\{G_{yx}^\text{H}(\omega)\right\} = \frac{\pi \omega \rho_0}{4} g_{g,1}^\text{H}(\omega),
\end{equation}
\begin{equation}
  \text{Im}\left\{G_{zz}^\text{H}(\omega)\right\} = \frac{\pi \omega \rho_0}{4}, g_{\parallel}^\text{H}(\omega)
\end{equation}
\end{subequations}

where $\rho_0 = \omega^2/\pi^2 c^3$ is the vacuum density of states and,

\begin{subequations}\label{Eq:g_m_anisotropic}
  \begin{equation}
  \begin{split}
    g_{\bot,1}^\text{H}(\omega) = \frac{1}{\pi} \int_{0}^{2\pi} d\phi \Bigg\{ & \int_{0}^{1} \frac{\kappa d\kappa }{p}\Big[ 1  + \sin^2\phi \text{Re}\left\{ r_{pp}e^{2ik_0 pd} \right\} - \kappa^2 \cos^2\phi + \cos^2\phi\left(\kappa^2 - 1\right)\text{Re}\left\{ r_{ss}e^{2ik_0 pd}\right\} \\
    + & p\sin\phi\cos\phi \text{Re}\left\{ e^{2ik_0 pd}(r_{ps} - r_{sp}) \right\} \Big] \\
    + & \int_{1}^{\infty} \frac{\kappa d\kappa}{|p|} \left[ \sin^2\phi \text{Im}\{ r_{pp} \} + \cos^2 \phi \left(\kappa^2 - 1\right) \text{Im}\{r_{ss}\} + |p|\sin\phi\cos\phi \text{Re}\left\{ r_{ps} - r_{sp} \right\} \right]e^{-2k_0 |p|d} \Bigg\},
  \end{split}
  \end{equation}
  \begin{equation}
  \begin{split}
    g_{\bot,2}^\text{H}(\omega) = \frac{1}{\pi} \int_{0}^{2\pi} d\phi \Bigg\{ & \int_{0}^{1} \frac{\kappa d\kappa }{p}\Big[ 1  + \cos^2\phi \text{Re}\left\{ r_{pp}e^{2ik_0 pd} \right\} - \kappa^2 \sin^2\phi + \sin^2\phi\left(\kappa^2 - 1\right)\text{Re}\left\{ r_{ss}e^{2ik_0 pd}\right\} \\
    - & p\sin\phi\cos\phi \text{Re}\left\{ e^{2ik_0 pd}(r_{ps} - r_{sp}) \right\} \Big] \\
    + & \int_{1}^{\infty} \frac{\kappa d\kappa}{|p|} \left[ \cos^2\phi \text{Im}\{ r_{pp} \} + \sin^2 \phi \left(\kappa^2 - 1\right) \text{Im}\{r_{ss}\} - |p|\sin\phi\cos\phi \text{Re}\left\{ r_{ps} - r_{sp} \right\} \right]e^{-2k_0 |p|d} \Bigg\},
  \end{split}
  \end{equation}
  \begin{equation}
    \begin{split}
       g_{g,1}^\text{H}(\omega) =  -\frac{1}{\pi} \int_{0}^{2\pi} \Bigg\{ & d\phi \int_{0}^{1} \kappa d\kappa \left[ \sin^2\phi \text{Im}\left\{ r_{ps} e^{2ik_0 pd} \right\} + \cos^2\phi \text{Im}\left\{ r_{sp} e^{2ik_0 pd} \right\}  \right] \\
       + & \int_{1}^{\infty} \kappa d\kappa \left[ \sin^2\phi \text{Im}\{r_{ps}\} + \cos^2\phi \text{Im}\{ r_{sp} \} \right]e^{-2k_0 |p|d}  \Bigg\},
    \end{split}
  \end{equation}
  \begin{equation}
    g_\parallel^\text{H}(\omega) = \frac{1}{2\pi} \int_{0}^{2\pi} d\phi \Bigg\{ \int_{0}^{1} \frac{\kappa^3 d\kappa}{p}\left( 1 + \text{Re}\left\{ r_{ss}e^{2ik_0 pd} \right\} \right) + \int_{1}^{\infty} \frac{\kappa^3 d\kappa}{|p|}e^{-2k_0 |p|d} \text{Im}\{r_{ss}\}  \Bigg\}.
  \end{equation}
\end{subequations}

Plugging Eq.~(\ref{Eq:G_tensor_components}) into Eq.~(\ref{Eq:P_mag_anisotropic_xy}), we find,
\begin{equation}
    P_\text{mag} = \int_{-\infty}^\infty d\omega \hbar \omega \Gamma^\text{H}(\omega),
\end{equation}
with,
\begin{equation}\label{Eq:Gamma_Rad_XY}
  \begin{split}
  \Gamma^\text{H}(\omega) = (\omega \rho_0/8) \Bigg\{ & \left[g_{\bot,1}^\text{H}(\omega) + g_{\bot,2}^\text{H}(\omega) + 2g_{g,1}^\text{H}(\omega) \right] \left[ \text{Im}\left\{\alpha_{m,\bot}(\omega^-)\right\} - \text{Re}\left\{ \alpha_{m,g}(\omega^-) \right\} \right] \left[n_1(\omega^-) - n_0(\omega)\right] \\
  + & 2 g_{\parallel}^\text{H}(\omega) \text{Im}\left\{ \alpha_{m,\parallel}(\omega) \right\}\left[n_1(\omega) - n_0(\omega)\right] \Bigg\}.
  \end{split}
\end{equation}

For the case when the YIG interface is the $x-z$ plane (Fig.~\ref{Fig:Schematic_XY_XZ}(b)), we find the radiated power by exchanging the axes $\hat{x} \to \hat{z}$, $\hat{y} \to \hat{x}$, and $\hat{z} \to \hat{y}$ in Eq.~(\ref{Eq:Greens_function_anisotropic}). In this case, we have
\begin{subequations}
  \begin{equation}\label{Eq:G_xx_XZplane}
    \text{Im}\left\{ G_{xx}^\text{H}(\omega) \right\} = \frac{\pi \omega \rho_0}{8} g_{\bot,2}^\text{H}(\omega),
  \end{equation}
  \begin{equation}\label{Eq:G_yy_XZplane}
    \text{Im}\left\{ G_{yy}^\text{H}(\omega) \right\} = \frac{\pi \omega \rho_0}{4} g_\parallel^\text{H}(\omega),
  \end{equation}
  \begin{equation}
    \text{Im}\left\{ G_{zz}^\text{H}(\omega) \right\} = \frac{\pi \omega \rho_0}{8} g_{\bot,1}^\text{H},
  \end{equation}
\end{subequations}
where $g_{\bot,1}^\text{H}$, $g_{\bot,2}^\text{H}$, and $g_{\parallel}^\text{H}$ given by Eq.~(\ref{Eq:g_m_anisotropic}). For the $xy$ and $yx$ component of the Green's function, however, we get
\begin{equation}\label{Eq:G_xy_re_XZplane}
  \text{Re}\left\{G_{xy}^\text{H}(\omega) \right\} - \text{Re}\left\{G_{yx}^\text{H}(\omega) \right\} = \frac{\pi\omega \rho_0}{4} g_{g,2}^\text{H} (\omega),
\end{equation}
with
\begin{equation}\label{Eq:g_g2_anisotropic}
\begin{split}
  g_{g,2}^\text{H}(\omega) = \frac{1}{\pi} \int_{0}^{2\pi}d\phi \Bigg\{ & \int_{0}^{1} \frac{\kappa^2 d\kappa}{p} \left[p \sin\phi \text{Im}\left\{r_{ss}e^{2ik_0 pd}\right\}  + \frac{\cos\phi}{2} \text{Im}\left\{ \left( r_{ps} - r_{sp}\right) e^{2ik_0 pd} \right\}  \right]  \\
  +  &  \int_{1}^{\infty} \frac{\kappa^2 \kappa}{|p|}\left[ |p|\sin\phi \text{Im}\{ r_{ss} \} - \frac{\cos\phi}{2} \text{Re}\left\{r_{sp} - r_{ps} \right\} \right] e^{-2k_0 |p|d} \Bigg\},
\end{split}
\end{equation}
and thus we have for the case when the YIG interface is the $x-z$ plane,
\begin{equation}\label{Eq:Gamma_Rad_XZ}
  \begin{split}
  \Gamma^\text{H}(\omega) = (\omega \rho_0/8) \Bigg\{ & \left[g_{\bot,2}^\text{H}(\omega) + 2g_{\parallel}^\text{H}(\omega) + 2g_{g,2}^\text{H}(\omega) \right] \left[ \text{Im}\left\{\alpha_{m,\bot}(\omega^-)\right\} - \text{Re}\left\{ \alpha_{m,g}(\omega^-) \right\} \right] \left[n_1(\omega^-) - n_0(\omega)\right] \\
  + & g_{\bot,1}^\text{H}(\omega) \text{Im}\left\{ \alpha_{m,\parallel}(\omega) \right\}\left[n_1(\omega) - n_0(\omega)\right] \Bigg\},
  \end{split}
\end{equation}
with $g_{\bot,1}^\text{H}$, $g_{\bot,2}^\text{H}$, and $g_{\parallel}^\text{H}$ given by Eq.~(\ref{Eq:g_m_anisotropic}) and $g_{g,2}^\text{H}$ by Eq.~(\ref{Eq:g_g2_anisotropic}). This is the same as Eq.~(\ref{Eq:Spetral_density_xz}) in the main manuscript. 

\section{Dominance of Magnetic Local Density of States}
Although the expressions found in the previous sections for the radiated power $P^\mathrm{rad}$ are not, in general, exactly proportional to the local density of states (LDOS), they are proportional to terms of the same order as the LDOS. The expression for LDOS is given by~\cite{joulain2003definition},
\begin{equation}
    \rho(\bm r, \omega) = \frac{1}{\pi \omega} \text{Tr}\left[ \epsilon_0 \text{Im}\left\{ G^\text{E}(\bm r, \bm r, \omega) \right\} + \text{Im}\left\{ G^\text{H}(\bm r, \bm r,\omega) \right\} \right],
\end{equation}
where the $\text{Tr}$ represents the trace  operator. Using the expressions of the previous section, it is easy to see that the LDOS at the location of the nanosphere is given by, 
\begin{equation}
    \rho(\omega) = (\rho_0/8)\left[ \epsilon_0 (g_{\bot,1}^\text{E} + g_{\bot,2}^\text{E} + 2 g_{\parallel}^\text{E}) + g_{\bot,1}^\text{H} + g_{\bot,2}^\text{H} + 2 g_{\parallel}^\text{H}  \right] ,
\end{equation}
where the expressions for $g_{\bot,1}^\text{H}$, $g_{\bot,2}^\text{H}$, and $g_{\parallel}^\text{H}$ are given by Eq.~(\ref{Eq:g_m_anisotropic}) and the expression for the electric Green's functions are found from the magnetic ones by replacing $s\to p$ and $p\to s$ and dividing by $\epsilon_0$. As discussed before, the magnetic Green's functions are about eight orders of magnitude larger than the electric ones at GHz frequencies, and thus, the LDOS is dominated by the magnetic LDOS. This shows that the magnetic field fluctuations dominate the vacuum radiation, vacuum torque, and LDOS simultaneously. 

\section{Magnetic Polarizability Tensor of YIG}\label{Sec:Polarizability_tensor}

In the appendix, we provide derivations of the YIG polarizability tensor. We consider the Landau-Lifshitz-Gilbert formula to describe the YIG permeability tensor~\cite{pozar2006microwave},
\begin{equation}\label{Eq:mu_matrix_YIG}
  \bar{\bar{\mu}} = \begin{pmatrix}
                      \mu_\bot & -\mu_g & 0 \\
                      \mu_g & \mu_\bot & 0 \\
                      0 & 0 & \mu_\parallel
                    \end{pmatrix},
\end{equation}
where
\begin{subequations}\label{Eq:mu_T_mu_g}
\begin{equation}
  \mu_\bot(\omega) = \mu_0 (1 + \chi_\bot) = \mu_0\left\{ 1+ \frac{\omega_0 \omega_m (\omega_0^2 - \omega^2) + \omega_0 \omega_m \omega^2 \alpha^2 + i \left\{ \alpha \omega \omega_m \left[ \omega_0^2 + \omega^2 (1 + \alpha^2) \right] \right\}}{\left[ \omega_0^2 - \omega^2 (1 + \alpha^2) \right]^2 + 4\omega_0^2 \omega^2 \alpha^2} \right\},
\end{equation}
\begin{equation}
  \mu_g(\omega) = \mu_0 \chi_g = \mu_0 \frac{-2\omega_0 \omega_m \omega^2 \alpha + i\omega \omega_m \left[ \omega_0^2 - \omega^2 (1+\alpha^2) \right]}{\left[ \omega_0^2 - \omega^2 (1 + \alpha^2) \right]^2 + 4\omega_0^2 \omega^2 \alpha^2},
\end{equation}
\begin{equation}
  \mu_\parallel = \mu_0,
\end{equation}
\end{subequations}
and $\omega_0 = \mu_0 \gamma H_0$ is the Larmor precession frequency with $\gamma$ being the gyromagnetic ratio and $H_0$ the bias magnetic field (assumed to be along $\hat{z}$ direction), $\omega_m = \mu_0 \gamma M_s$ with $M_s$ being the saturation magnetization of the material, and $\alpha$ is the YIG damping factor related to the width of the magnetic resonance through $\Delta H = 2\alpha \omega / \mu_0 \gamma$. In the main text, we considered $M_s=1780\,\mathrm{Oe}$ and $\Delta H=45\,\mathrm{Oe}$~\cite{pozar2006microwave} in our calculations.

When the magnetic field is reversed (along $-\hat{z}$ direction), we can use the same results by doing the substitutions
\begin{equation}
  \omega_0 \to -\omega_0, \quad \omega_m \to -\omega_m, \quad \alpha \to -\alpha,
\end{equation}
which gives
\begin{equation}
  \mu_\bot \to \mu_\bot, \quad \mu_g \to -\mu_g.
\end{equation}

Using the method in Ref.~\cite{Sihvola1994dielectric} for the polarizability tensor of a sphere with arbitrary anisotropy, we find the polarizability tensor of YIG with the permeability tensor described by Eq.~(\ref{Eq:mu_matrix_YIG}),
\begin{equation}\label{Eq:alpha_H_anisotropic}
\begin{split}
  \bar{\bar{\alpha}}_\text{m} & = 4\pi a^3
                        \begin{pmatrix}
                         \frac{(\mu_\bot - \mu_0)(\mu_\bot + 2\mu_0) + \mu_g^2 }{(\mu_\bot + 2\mu_0)(\mu_\bot + 2\mu_0) + \mu_g^2} & \frac{-3\mu_0 \mu_g}{(\mu_\bot + 2\mu_0)(\mu_\bot + 2\mu_0) + \mu_g^2}  & 0 \\
                         \frac{3 \mu_0 \mu_g}{(\mu_\bot + 2\mu_0)(\mu_\bot + 2\mu_0) + \mu_g^2} & \frac{(\mu_\bot - \mu_0)(\mu_\bot + 2\mu_0)+\mu_g^2}{(\mu_\bot + 2\mu_0)(\mu_\bot + 2\mu_0) + \mu_g^2} & 0 \\
                         0 & 0 & \frac{\mu_\parallel - \mu_0}{\mu_z + 2\mu_0}
                       \end{pmatrix}.
\end{split}
\end{equation}
Therefore the magnetic polarizability terms in Eqs.~(\ref{Eq:Gamma_Rad_XY}) and (\ref{Eq:Gamma_Rad_XZ}) are given by,
\begin{subequations}
\begin{equation}
    \alpha_{m,\bot}(\omega) = 4\pi a^3 \frac{(\mu_\bot - \mu_0)(\mu_\bot + 2\mu_0) + \mu_g^2 }{(\mu_\bot + 2\mu_0)(\mu_\bot + 2\mu_0) + \mu_g^2},
\end{equation}
\begin{equation}
    \alpha_{m,g}(\omega) = 4\pi a^3 \frac{3\mu_0 \mu_g}{(\mu_\bot + 2\mu_0)(\mu_\bot + 2\mu_0) + \mu_g^2},
\end{equation}
\end{subequations}
where $\mu_\bot$ and $\mu_g$ are frequency dependent terms give by Eq.~(\ref{Eq:mu_T_mu_g}). 

It is important to note that magnetostatic approximation has been assumed in the derivation of the magnetic polarizability. This is similar to the electrostatic approximation used for the derivation of the electric polarizability \cite{bohren2008absorption}, where, using the duality of electromagnetic theory, the electric fields and electric dipoles have been replaced by the magnetic fields and magnetic dipoles. In this approximation, the fields inside the sphere are assumed to be constant. 

One can apply the Mie theory to find the magnetic polarizability to the first order in the Mie scattering components. This, however, is mathematically challenging due to the anisotropy of the magnetic material. For the purpose of our study, the magnetostatic assumption is enough to find the polarizability properties of YIG since the size of the sphere is much smaller compared to the wavelength, and the polarizability is dominated by the magneto-static term. 

For metals, however, higher order terms are important for finding the magnetic polarizability since the magneto-static terms are zero and only higher order terms due to electric dipole fluctuations give rise to the magnetic polarizability of metals \cite{manjavacas2010thermal}. We provide derivations based on Mie theory for the polarizability constant of an aluminum particle in Section S1 in the supplementary material. 

\section{Barnett Effect}\label{Sec:barnett}
In the simplest models of magnetic materials, electrons are assumed to be magnetic dipoles with the moments $\mu_B$ spinning about the magnetization axis determined by the applied magnetic field $H_0$ with the Larmor precession frequency $\omega_0 = \mu_0 \gamma H_0$, where $\gamma$ is the gyromagnetic ratio of the material \cite{pozar2006microwave}. Barnett showed that the spontaneous magnetization of a material with the magnetic susceptibility of $\chi$ is given by \cite{barnett1915magnetization}
\begin{equation}
    \bm M_{\text{rot}} = \chi \Omega/\gamma,
\end{equation}
where $\Omega$ is the rotation frequency of the magnetic material. This magnetization can be assumed to be caused by an applied magnetic field $H_\text{rot}$ which is $H_\text{rot} = \bm M_\text{rot}/\chi = \frac{\Omega}{\gamma \mu_0}$. We thus get the Larmor frequency due to rotation,
\begin{equation}
    \omega_{0,\text{rot}} = \Omega.
\end{equation}
Therefore, the Larmor frequency of a spinning magnetic material is the same as the rotation frequency. We thus can write the total Larmor frequency of spinning YIG as 
\begin{equation}
    \omega_0 = \Omega + \mu_0 \gamma H_0.
\end{equation}
We use this expression to find the permeability tensor of a spinning YIG nanosphere discussed in Appendix~\ref{Sec:Polarizability_tensor}. 

\section{Non-local Model for Aluminum}\label{Sec:nonlocalal}
Since the sphere is spinning in close proximity to material interfaces, the non-local effects in aluminum electromagnetic response can become important. Here, we employ the non-local Fresnel reflection coefficients from Ref.~\cite{ford1984electromagnetic}. 
\begin{equation}
    r_{ss} = \frac{Z^s - \frac{4\pi}{c p}}{Z^s + \frac{4\pi}{c p}}, \quad r_{pp} = \frac{4\pi p/c - Z^p}{4\pi p/c + Z^p},
\end{equation}
where $p = \sqrt{1 - \kappa^2 }$, and
\begin{subequations}
  \begin{equation}
    Z^s = \frac{8i}{c} \int_{0}^{\infty} dq \frac{1}{\epsilon_t(k,\omega) - (q^2 + \kappa^2)},
  \end{equation}
  \begin{equation}
    Z^p = \frac{8i}{c} \int_{0}^{\infty} dq \frac{1}{q^2 + \kappa^2} \left( \frac{q^2}{\epsilon_t(k,\omega) - (q^2 + \kappa^2)} + \frac{\kappa^2}{\epsilon_l(k,\omega)} \right),
  \end{equation}
\end{subequations}
with the longitudinal and transverse dielectric permittivities given by
\begin{subequations}
  \begin{equation}
    \epsilon_l(k,\omega) = 1 + \frac{3\omega_p^2}{k^2 v_F^2} \frac{(\omega + i\Gamma)f_l(u)}{ \omega + i\Gamma f_l(u)},
  \end{equation}
  \begin{equation}
    \epsilon_t(k,\omega) = 1 - \frac{\omega_p^2}{\omega(\omega + i\Gamma)} f_t(u),
  \end{equation}
\end{subequations}
with $k^2 = (\omega/c)^2 \left(q^2 + \kappa^2 \right)$, $u = (\omega + i\Gamma)/(kv_F)$, and
\begin{equation}
  f_l(u) = 1 - \frac{1}{2}u \ln\frac{u+1}{u-1},\quad f_t(u) = \frac{3}{2}u^2 - \frac{3}{2}u(u^2 - 1)\ln \frac{u+1 }{u - 1}.
\end{equation}
These expressions give the non-local reflection coefficients at a metallic interface for the semi-classical infinite barrier (SCIB) model. %Note that $r_{ss}$ and $r_{pp}^p$ are the same as the $r_{ss}$ and $r_{pp}$ introduced in the previous section. 
The SCIB model is accurate as long as $z = \frac{k}{2k_F} \sim 0$, where $k_F = m v_F/\hbar$ with $m$ being the free-electron mass. For example, for aluminum with $v_F \simeq 2.03\times 10^6 \text{m}/\text{s}$, we have $k_F \simeq 1.754\times 10^{10}$ and $k = \omega/c \simeq 20$, which shows that for our case the SCIB model is valid.

\section{Vacuum Frictional Torque}\label{Sec:torque}

In this section, we provide the derivations of the vacuum frictional torque exerted on the spinning YIG nanosphere due to vacuum fluctuations. The torque on a magnetic dipole is given by
\begin{equation}
    \bm M = \bm m\times \bm H.
\end{equation}
Since we are interested in the torque along the rotation axis ($z$ direction), we can write the torque as
\begin{equation}
\begin{split}
  M_z = & \hat{z}\cdot \langle \bm m^\text{fl}\times \bm H^\text{ind} + \bm m^\text{ind}\times \bm H^\text{fl} \rangle \\
  & = \langle m_x^\text{fl} H_y^\text{ind} - m_y^\text{fl} H_x^\text{ind} + m_x^\text{ind} H_y ^\text{fl} - m_y^\text{ind} H_x^\text{fl} \rangle,
\end{split}
\end{equation}
using the Fourier transform, we get
\begin{equation}
  \begin{split}
     M_z = \int \frac{d\omega d\omega'}{(2\pi)^2} & e^{-i(\omega + \omega')t}\Big[ \langle m_x^\text{fl}(\omega) H_y^\text{ind}(\omega') \rangle - \langle m_y^\text{fl}(\omega) H_x^\text{ind}(\omega') \rangle + \langle m_x^\text{ind}(\omega) H_y^\text{fl}(\omega')\rangle - \langle m_y^\text{ind}(\omega) H_x^\text{fl}(\omega')\rangle \Big].
  \end{split}
\end{equation}
Through a similar approach as that used in Appendix~\ref{Sec:Vacuum_Radiation}, after some algebra, we find 
\begin{equation}
  \begin{split}
     M_z = \frac{\hbar}{2\pi} \int_{-\infty}^{\infty} d\omega \Bigg\{ & \left( \text{Im}\left\{G_{yy}^\text{H}(\omega)\right\} + \text{Im}\left\{ G_{xx}^\text{H}(\omega) \right\} + \text{Re} \left\{ G_{yx}^\text{H}(\omega)\right\} - \text{Re} \left\{G_{xy}^\text{H}(\omega)\right\} \right)\times  \\
     & \left[ \text{Im}\left\{ \alpha_{m,\bot}(\omega^+)\right\} + \text{Re}\left\{ \alpha_{m,g}(\omega^+) \right\} \right]\left[ n_1(\omega^+) - n_0(\omega) \right] \\
     - & \left( \text{Im}\left\{ G_{yy}^\text{H}(\omega) \right\} + \text{Im} \left\{ G_{xx}^\text{H}(\omega) \right\} - \text{Re} \left\{ G_{yx}^\text{H}(\omega)\right\} + \text{Re} \left\{ G_{xy}^\text{H}(\omega)\right\} \right)\times  \\
     & \left[\text{Im}\left\{ \alpha_{m,\bot}(\omega^-) \right\} - \text{Re}\left\{ \alpha_{m,g}(\omega^-) \right\} \right]\left[n_1(\omega^-) - n_0(\omega)\right] \Bigg\},
  \end{split}
\end{equation}
which can be written as
\begin{equation}\label{Eq:Mz_01}
    M_z = - \int_{-\infty}^{+\infty} d\omega \hbar \Gamma^\text{H}_\text{M}(\omega).
\end{equation}

For an interface in the $x-y$ plane $\Gamma^\text{H}_M$ is given by
\begin{equation}
  \Gamma^\text{H}_M (\omega) = (\omega \rho_0/8)\left[ g_{\bot,1}^\text{H}(\omega) + g_{\bot,2}^\text{H}(\omega) + 2 g_{g,1}^\text{H}(\omega) \right]\left[\text{Im}\left\{ \alpha_{m,\bot}(\omega^-) \right\} - \text{Re}\left\{ \alpha_{m,g}(\omega^-) \right\} \right]\left[n_1(\omega^-) - n_0(\omega)\right],
\end{equation}
which is the same expression for the radiated power minus the term related to the axis of rotation $z$. For an interface in the $x-z$ plane, on the other hand, $\Gamma^\text{H}_M$ we have
\begin{equation}
  \Gamma^\text{H}_M (\omega) = (\omega \rho_0/8)\left[ g_{\bot,2}^\text{H}(\omega) + 2g_{\parallel}^\text{H}(\omega) + 2 g_{g,2}^\text{H}(\omega) \right]\left[\text{Im}\left\{ \alpha_{m,\bot}(\omega^-) \right\} - \text{Re}\left\{ \alpha_{m,g}(\omega^-) \right\} \right]\left[n_1(\omega^-) - n_0(\omega)\right].
\end{equation}
This expression is the same as Eq.~(\ref{Eq:Spetral_density_xz}) in the main manuscript, with the difference that it does not have the last term involving the term $n_1(\omega) - n_0(\omega)$. Compared to the vacuum radiation expression, vacuum torque has an extra minus sign in Eq.~(\ref{Eq:Mz_01}), indicating that this torque acts as friction rather than a driving force, as expected. 

\subsection{Other components of torque}
In the previous section, we only derived the $z$ components of the torque exerted on the nanosphere. The $x$ and $y$ components can be written as
\begin{subequations}
  \begin{equation}
    M_x = \langle m_y^\text{fl} H_z^\text{ind} - m_z^\text{fl} H_y^\text{ind} + m_y^\text{ind} H_z^\text{fl} - m_z^\text{ind} H_y^\text{fl} \rangle,
  \end{equation}
  \begin{equation}
    M_y = \langle m_z^\text{fl} H_x^\text{ind} - m_x^\text{fl} H_z^\text{ind} + m_z^\text{ind} H_x^\text{fl} - m_x^\text{ind} H_z^\text{fl} \rangle.  \end{equation}
\end{subequations}
Using a similar approach as that used in the previous section and section \ref{Sec:Vacuum_Radiation}, incorporating the torque due to the electric field fluctuations of vacuum and the magnetic dipole fluctuations of the YIG sphere, we find for the $x$ component of torque,
\begin{equation}\label{Eq:Mx_01}
  \begin{split}
     M_x = \frac{\hbar}{4\pi} \int_{-\infty}^{\infty} d\omega \Bigg\{ & \left[2n_1(\omega^-) + 1\right] \left[ \text{Im}\left\{ \alpha_{m,\bot}(\omega^-) \right\} - \text{Re}\left\{ \alpha_{m,g}(\omega^-) \right\} \right]\left[2\text{Im}\left\{ G_{zx}^\text{H}(\omega) \right\} + 2\text{Re} \left\{G_{zy}^\text{H}(\omega) \right\} \right] \\
     - & 4\left[ n_1(\omega) + 1 \right] \text{Im}\left\{ \alpha_{m,\parallel}(\omega) \right\} \text{Re}\left\{G_{yz}^\text{H}(\omega) \right\} \\
     + \left[2 n_0(\omega) + 1\right] \Bigg\{ & \left[ \text{Re}\{\alpha_{m,\bot}(\omega^-)\} + \text{Im}\{\alpha_{m,g}(\omega^-)\} \right] \left(\text{Re}\{G_{xz}^\text{H}(\omega)\} - \text{Re} \{G_{zx}^{\text{H}}(\omega)\} + \text{Im} \{G_{yz}^\text{H}(\omega)\} + \text{Im} \{ G_{zy}^{\text{H}}(\omega)\} \right) \\
     + & \left[ \text{Im}\{\alpha_{m,\bot}(\omega^-)\} - \text{Re}\{\alpha_{m,g}(\omega^-)\} \right] \left(-\text{Im}\{G_{xz}^\text{H}(\omega)\} - \text{Im} \{G_{zx}^{\text{H}}(\omega)\} + \text{Re} \{G_{yz}^\text{H}(\omega)\} - \text{Re} \{ G_{zy}^{\text{H}}(\omega)\} \right) \Bigg\} \\
     + \left[n_0(\omega) + 1\right] \Bigg\{- & 2\text{Re}\{\alpha_{m,\parallel}(\omega)\}\left( \text{Im}\{ G_{zy}^\text{H}(\omega)\} + \text{Im} \{G_{yz}^{\text{H}}(\omega)\} \right) + 2\text{Im}\{\alpha_{m,\parallel}(\omega)\}\left( -\text{Re}\{ G_{zy}^\text{H}(\omega)\} + \text{Re} \{G_{yz}^{\text{H}}(\omega)\} \right)  \Bigg\},\\
  \end{split}
\end{equation}
and for the $y$ component,
\begin{equation}\label{Eq:My_01}
\begin{split}
  M_y = \frac{\hbar}{4\pi} \int_{-\infty}^{\infty} d\omega \Bigg\{ & \left[2n_1(\omega^-) + 1\right]\left[ \text{Im}\left\{ \alpha_{m,\bot}(\omega^-) \right\} - \text{Re}\left\{ \alpha_{m,g}(\omega^-) \right\}\right]\left[ -2\text{Re}\{G_{zx}^\text{H}(\omega)\} + 2\text{Im}\{G_{zy}^\text{H}(\omega)\}  \right]  \\
  + & 4 \left[ n_1(\omega) + 1 \right] \text{Im}\left\{ \alpha_{m,\parallel}(\omega) \right\} \text{Re}\{G_{xz}^\text{H}(\omega)\} \Bigg\} \\
  - \left[ 2n_0(\omega) + 1\right] \Bigg\{ & \left[ \text{Re}\{\alpha_{m,\bot}(\omega^-)\} + \text{Im}\{\alpha_{m,g}(\omega^-)\} \right]\left( \text{Im}\{G_{xz}^\text{H}(\omega)\} + \text{Im}\{G_{zx}^{\text{H}}(\omega)\} - \text{Re}\{G_{yz}^\text{H}(\omega)\} + \text{Re}\{ G_{zy}^{\text{H}}(\omega)\} \right) \\
  + & \left[ \text{Im}\{\alpha_{m,\bot}(\omega^-)\} - \text{Re}\{\alpha_{m,g}(\omega^-)\} \right]\left( \text{Re}\{G_{xz}^\text{H}(\omega)\} - \text{Re}\{G_{zx}^{\text{H}}(\omega)\} + \text{Im}\{G_{yz}^\text{H}(\omega)\} + \text{Im}\{ G_{zy}^{\text{H}}(\omega)\} \right) \Bigg\} \\
  -\left[ n_0(\omega) + 1 \right] & \Bigg\{-2\text{Re}\alpha_{m,\parallel}(\omega) \left( \text{Im}\{G_{zx}^\text{H}(\omega)\} + \text{Im}\{G_{xz}^{\text{H}}(\omega)\} \right) + 2\text{Im}\alpha_{m,\parallel}(\omega) \left( -\text{Re}\{G_{zx}^\text{H}(\omega)\} + \text{Re}\{G_{xz}^{\text{H}}(\omega)\} \right) \Bigg\}.
  \end{split}
\end{equation}
We can find the $x$ and $y$ components of frictional torque by plugging magnetic Green's function expressions into Eqs.~(\ref{Eq:Mx_01}) and (\ref{Eq:My_01}). 

Remarkably, we find that the spinning YIG nanosphere can experience a large torque along the x or y direction when the YIG interface is biased by external magnetic fields in the x or y direction. This means that in these cases, the sphere can rotate out of the rotation axis and start to precess. This will change the validity of the equations found for the vacuum radiation and frictional torque along the $z$ axis since it has been assumed that the sphere is always rotating around the $z$ axis and is also magnetized along that axis. However, this torque is still small enough compared to the driving torque of the trapping laser and it will still give enough time to make the observations of vacuum fluctuation effects. In Section S2 in the supplementary material, we present the plots of these torques when the interface is the $x-y$ or $x-z$ plane and provide more detailed discussions.

\subsection{Recoil torque}
Another contribution to the torque comes from the case when the induced dipole moments on the YIG sphere re-radiate due to the vacuum electric field fluctuations. This causes a recoil torque on the sphere and can be written as
\begin{equation}
  \bm M^\text{rec} = \langle \bm m^\text{ind} \times \bm H^\text{sc} \rangle,
\end{equation}
where $\bm H^\text{sc}$ is the scattered fields from the dipole and are given by,
\begin{equation}
  \bm H^\text{sc}(\bm r, \omega) = \bar{\bm G}^\text{H}(\bm r,\bm r',\omega) \cdot \bm m^\text{ind}(\bm r',\omega),
\end{equation}
which shows that this term is of higher order contribution. We find that this recoil torque is much smaller than the torque derived in Eq.~(\ref{Eq:Mz_01}) for YIG spheres spinning near YIG or Al interfaces and can thus be ignored in all studied cases. We provide detailed derivations of $\bm M^\text{rec}$ and quantitative comparisons in Section S2 in the supplementary material.

\section{Experimental Analysis}\label{Sec:exptana}

In this section, we present the analytical steps for finding the experimental prediction plots provided in the last section of the main text. 

\subsection{Effects of drag torque due to imperfect vacuum}
In the real system of a spinning sphere, the environment is not a pure vacuum. This causes an extra torque on the spinning sphere from air molecules in the imperfect vacuum. The steady-state spin of the sphere happens when the driving torque of the trapping laser is equal to the drag and vacuum friction torques. In the case when there is no interface present, the only important counteracting torque is the drag torque given by~\cite{corson2017calculating}
\begin{equation}
  M_{\text{drag}} = \frac{2\pi \mu a^4}{1.497 \lambda} \Omega,
\end{equation}
where $a$ is the sphere radius, $\mu$ is the viscosity of the gas the sphere is spinning in, $\lambda$ is the mean free path of the air molecules, and $\Omega$ is the rotation frequency. We further have for gases \cite{kruger1965introduction},
\begin{equation}
  \lambda  = \frac{\mu}{p_{\text{gas}}} \sqrt{\frac{\pi K_\text{B}T}{2m}},
\end{equation}
where $p_\text{gas}$ and $m$ are the pressure and the molecular mass of the gas, respectively. Thus, we get the drag torque,
\begin{equation}
  M_\text{drag} = \frac{2 a^4 p_\text{gas}}{1.479}\sqrt{\frac{2\pi m}{k_B T}} \Omega.
\end{equation}
For 1 GHz rotation of a sphere, the balance between the drag torque and the optical torque $M_\text{opt}$ happens at about $p_\text{gas} = 10^{-4}$ torr. Therefore we get, at room temperature and for a molecular mass of $28.966 \text{gram}/\text{mol}$,
\begin{equation}
  \sqrt{\frac{2\pi m}{K_B T}} = 8.542\times 10^{-3},
\end{equation}
and thus \cite{ahn2018optically},
\begin{equation}
  M_\text{opt} = 1.568\times 10^{-21} \text{N}\cdot \text{m},
\end{equation}
This is important for studying the effects of vacuum torque on the rotation speed of the sphere. As shown in the main text, we find that for vacuum pressures of about $10^{-4}$ torr, changes in the balance speed of the YIG nanoparticle when it is closer to material interfaces are detectible in the power spectral density (PSD) of the nanosphere~\cite{ahn2018optically}.

\subsection{Effects of negative torque and shot noise heating due to the trapping laser}
When the trapping laser is linearly polarized, it can exert a negative torque on the spinning particle. The torque on the sphere due to the laser is given by $\bm M_\text{opt} = \frac{1}{2} \text{Re}\{ \bm p^* \times \bm E \}$~\cite{ahn2018optically}, where $\bm p$ is the dipole moment of the sphere, given by $\bm p = \bar{\bm \alpha}^{\text{eff}}\cdot \bm E$, with $\bar{\bm \alpha}^\text{eff}$ being the effective polarizability of the sphere as seen in the frame of the lab, and $\bm{E}$ is the electric field from the laser. As derived in Section S3 in the supplementary material, in the case when the laser is linearly polarized, the negative torque from the laser is proportional to $\text{Im}\{\alpha(\omega_0 + \Omega)\} - \text{Im}\{\alpha(\omega_0 - \Omega)\}$, where $\omega_0 = 1.21\times 10^{16}$ is the frequency of the laser, and $\Omega = 6.28\times 10^9$ is the rotation frequency. Since $\Omega \ll \omega_0$, we get $\alpha(\omega^+) \simeq \alpha(\omega^-)$ and thus the second term is negligible. We can thus ignore the negative torque coming from the laser when the laser is linearly polarized.

Another effect from the trapping laser is the heating of nanoparticles due to the shot noise. The rate of temperature change due to shot noise heating can be determined by the laser frequency, the power of the laser per unit area, the mass of the particle, and the scattering cross section for the nanoparticles~\cite{seberson2019distribution}. For YIG nanospheres of density $5110 \text{kg}/\text{m}^3$ and radius $200$ nm, and trapping laser of $1550$ nm wavelength and of $500$ mW power focused on an area of radius $0.7566 \mu$m, we find that the temperature change due to shot noise is small compared to the time scale of the rotation, which is $1$ ns. Therefore, the thermodynamic equilibrium condition for the FDT is valid. We provide further details for the derivations and calculations of negative torque and shot noise heating due to the trapping laser in Section S3 in the supplementary material.

\twocolumngrid
% \Urlmuskip=0mu plus 1mu\relax
% \bibliographystyle{apsrev4-2}
\bibliography{main}% Produces the bibliography via BibTeX.
%TC:endignore

\clearpage



\section*{Supplementary material}

# SUPPLEMENTAL MATERIAL FOR 'GIANT ENHANCEMENT OF VACUUM FRICTION IN SPINNING YIG NANOSPHERES'

**Farhad Khosravi[1,2], Wenbo Sun[2], Chinmay Khandekar[2], Tongcang Li[2,3], and Zubin Jacob[2,*]**

[1]Department of Electrical and Computer Engineering, University of Alberta,
Edmonton, Alberta T6G 1H9, Canada
[2]Elmore Family School of Electrical and Computer Engineering,
Birck Nanotechnology Center, Purdue University, West Lafayette, Indiana 47907, USA
[3]Department of Physics and Astronomy, Purdue Quantum Science and Engineering Institute,
Purdue University, West Lafayette, Indiana 47907, USA
[*]zjacob@purdue.edu

## Contents

## S1   Non-Electrostatic Limit and Magnetic Polarizability due to Electric Fluctuations

In this section, we provide derivations for the magnetic polarizability of metallic nanoparticles due to the electric dipole terms based on Mie theory. If a sphere is placed in the direction of a plane wave polarized along $\hat{x}$ direction and propagating along $z$ direction

$$\boldsymbol{E}_i = E_0 e^{ik_0 r \cos\theta} \hat{x}, \tag{S1}$$

The scattered fields are given by [2],

$$\boldsymbol{E}_s = -\sum_{n=1}^{\infty} E_n \left( i a_n \boldsymbol{N}_{e1n}^{(1)} - b_n \boldsymbol{M}_{o1n}^{(1)} \right), \tag{S2}$$

$$\boldsymbol{H}_s = -\frac{k_0}{\omega\mu} \sum_{n=1}^{\infty} E_n \left( i b_n \boldsymbol{N}_{o1m}^{(1)} + a_n \boldsymbol{M}_{e1n}^{(1)} \right), \tag{S3}$$

where

$$\boldsymbol{M}_{emn} = \frac{-m}{\sin\theta} \sin m\phi P_n^m(\cos\theta) z_n(kr)\hat{\theta} - \cos m\phi \frac{dP_n^m(\cos\theta)}{d\theta} z_n(kr)\hat{\phi}, \tag{S4a}$$

$$\boldsymbol{M}_{omn} = \frac{m}{\sin\theta} \cos m\phi P_n^m(\cos\theta) z_n(kr)\hat{\theta} - \sin m\phi \frac{dP_n^m(\cos\theta)}{d\theta} z_n(kr)\hat{\phi}, \tag{S4b}$$

$$\boldsymbol{N}_{emn} = \frac{z_n(kr)}{kr} \cos m\phi n(n+1) P_n^m(\cos\theta)\hat{r} + \cos m\phi \frac{dP_n^m(\cos\theta)}{d\theta} \frac{1}{kr} \frac{d}{d(kr)} \left[ krz_n(kr) \right] \hat{\theta}$$
$$- m \sin m\phi \frac{P_n^m(\cos\theta)}{\sin\theta} \frac{1}{kr} \frac{d}{d(kr)} \left[ krz_n(kr) \right] \hat{\phi}, \tag{S4c}$$

$$\boldsymbol{N}_{omn} = \frac{z_n(kr)}{kr}\sin m\phi\, n(n+1) P_n^m(\cos\theta)\hat{r} + \sin m\phi \frac{dP_n^m(\cos\theta)}{d\theta}\frac{1}{kr}\frac{d}{d(kr)}\left[krz_n(kr)\right]\hat{\theta}$$
$$+ m\cos m\phi \frac{P_n^m(\cos\theta)}{\sin\theta}\frac{1}{kr}\frac{d}{d(kr)}\left[krz_n(kr)\right]\hat{\phi}, \tag{S4d}$$

the superscripts (1) for $\boldsymbol{M}$ and $\boldsymbol{N}$ indicate that the Bessel functions are the Hankel functions of the first kind $h^{(1)}(kr)$, $E_n = i^n E_0(2n+1)/n(n+1)$, and $a_n$ and $b_n$ are the Mie scattering coefficients. On the other hand, the radiated fields due to an electric dipole are given by

$$\boldsymbol{E}_d = \frac{k_0^3}{4\pi\epsilon_m}\left\{(\hat{r}\times\boldsymbol{p})\times\hat{r}\frac{e^{ikr}}{kr} + [3\hat{r}(\hat{r}\cdot\boldsymbol{p}) - \boldsymbol{p}]\left(\frac{1}{(kr)^3} - \frac{i}{(kr)^2}\right)e^{ikr}\right\}, \tag{S5a}$$

$$\boldsymbol{H}_d = \frac{ck_0^2}{4\pi}(\hat{r}\times\boldsymbol{p})\frac{e^{ikr}}{r}\left(1 - \frac{1}{ikr}\right). \tag{S5b}$$

Using the facts that

$$P_1^1(\cos\theta) = -\sin\theta, \quad \frac{dP_1^1(\cos\theta)}{d\theta} = -\cos\theta, \tag{S6}$$

$$h_1^{(1)}(kr) = -e^{ikr}\left(\frac{i}{(kr)^2} + \frac{1}{kr}\right), \quad \frac{1}{kr}\frac{d}{d(kr)}\left[krh_1^{(1)}(kr)\right] = -e^{ikr}\left(-\frac{i}{(kr)^3} - \frac{1}{(kr)^2} + \frac{i}{kr}\right). \tag{S7}$$

The scattered fields to the first order of $n$ become

$$\boldsymbol{E}_s = \frac{3}{2}E_0\left\{ia_1\left[e^{ikr}\left(\frac{1}{(kr)^3} - \frac{i}{(kr)^2}\right)2\cos\phi\sin\theta\hat{r} - (\cos\phi\cos\theta\hat{\theta} - \sin\phi\hat{\phi})e^{ikr}\left(\frac{1}{(kr)^3} - \frac{i}{(kr)^2} - \frac{1}{kr}\right)\right]\right.$$
$$\left. - b_1\left[(\cos\phi\hat{\theta} - \sin\phi\cos\theta\hat{\phi})e^{ikr}\left(\frac{-1}{(kr)^2} + \frac{i}{kr}\right)\right]\right\}. \tag{S8}$$

Assuming that the dipole is along $x$ direction $\boldsymbol{p} = p_0\hat{x}$, the dipole fields become

$$\boldsymbol{E}_d = \frac{p_0 k_0^3}{4\pi\epsilon_m}\left\{(\cos\theta\cos\phi\hat{\theta} - \sin\phi\hat{\phi})\frac{e^{ikr}}{kr} + (2\hat{r}\sin\theta\cos\phi - \hat{\theta}\cos\theta\cos\phi + \hat{\phi}\sin\phi)\left(\frac{1}{(kr)^3} - \frac{i}{(kr)^2}\right)e^{ikr}\right\}$$
$$= \frac{p_0 k_0^3}{4\pi\epsilon_m}\left\{2\hat{r}\sin\theta\cos\phi\left(\frac{1}{(kr)^3} - \frac{i}{(kr)^2}\right)e^{ikr} - (\hat{\theta}\cos\theta\cos\phi - \hat{\phi}\sin\phi)\left(\frac{1}{(kr)^3} - \frac{i}{(kr)^2} - \frac{1}{kr}\right)e^{ikr}\right\}. \tag{S9}$$

In the low-frequency limit when $kr = \frac{2\pi r}{\lambda} \ll 1$, the scattered fields are dominated by terms of the order $(kr)^{-3}$. Thus, we can neglect the contribution from the $\boldsymbol{M}$ terms or the $b_1$ terms in Eq. (S8). In this limit, the fields of the dipole and the scattered fields become equivalent, if we take

$$p_0 = \frac{6\pi\epsilon_m i a_1}{k_0^3}E_0, \tag{S10}$$

or in other words, the sphere takes the polarizability

$$\alpha_e = \frac{6\pi\epsilon_m c^3}{\omega^3}ia_1, \tag{S11}$$

where

$$a_n = \frac{\epsilon_1 j_n(x_1)[x_0 j_n(x_0)]' - \epsilon_0 j_n(x_0)[x_1 j_n(x_1)]'}{\epsilon_1 j_n(x_1)[x_0 h_n^{(1)}(x_0)]' - \epsilon_0 h_n^{(1)}(x_0)[x_1 j_n(x_1)]'}, \tag{S12}$$

with $x_0 = k_0 a$, $x_1 = k_1 a$, and $k_1 = \omega\sqrt{\mu_1\epsilon_1}$, and $\mu_1$ and $\epsilon_1$ being properties of the sphere.

Now, we look at the scattered magnetic fields. We have to the first order

$$\boldsymbol{H}_s = \frac{3}{2}\frac{k_0}{\omega\mu_0}E_0\left\{ib_1\left[2\hat{r}\sin\phi\sin\theta\left(\frac{1}{(kr)^3} - \frac{i}{(kr)^2}\right)e^{ikr} - (\hat{\theta}\sin\phi\cos\theta + \hat{\phi}\cos\phi)\left(\frac{1}{(kr)^3} - \frac{i}{(kr)^2} - \frac{i}{kr}\right)e^{ikr}\right]\right.$$
$$\left. - a_n\left[(\hat{\phi}\cos\theta\cos\phi + \hat{\theta}\sin\phi)\left(\frac{i}{(kr)^2} + \frac{1}{kr}\right)e^{ikr}\right]\right\}. \tag{S13}$$

Again, we can ignore the second line or, in other words, $a_n$ in this expression for low frequencies. Then, comparing this expression with the magnetic fields of a magnetic dipole polarized along $\hat{y}$ direction $\boldsymbol{m} = m_0 \hat{y}$,

$$\boldsymbol{H}_m = \frac{m_0 k_0^3}{4\pi} \left\{ 2\hat{r} \sin\theta \sin\phi \left( \frac{1}{(kr)^3} - \frac{i}{(kr)^2} \right) e^{ikr} - (\hat{\theta} \cos\theta \sin\phi + \hat{\phi} \cos\phi) \left( \frac{1}{(kr)^3} - \frac{i}{(kr)^2} - \frac{1}{kr} \right) e^{ikr} \right\},$$
(S14)

Taking $H_0 = \frac{k_0}{\omega \mu_0} E_0$, we find that the two are equivalent if we have

$$m_0 = \frac{6\pi i b_1}{k_0^3} H_0,$$
(S15)

or if the sphere takes the magnetic polarizability

$$\alpha_m = \frac{6\pi c^3}{\omega^3} i b_1,$$
(S16)

where

$$b_n = \frac{\mu_1 j_n(x_1)[x_0 j_n(x_0)]' - \mu_0 j_n(x_0)[x_1 j_n(x_1)]'}{\mu_1 j_n(x_1)[x_0 h_n^{(1)}(x_0)]' - \mu_0 h_n^{(1)}(x_0)[x_1 j_n(x_1)]'}.$$
(S17)

In the low-frequency limit, we have

$$\lim_{x \to 0} j_n(x) = \frac{2^n n!}{(2n+1)!} x^n,$$
(S18)

and

$$\lim_{x \to 0} y_n(x) = -\frac{(2n)!}{2^n n!} \frac{1}{x^{n+1}}.$$
(S19)

Therefore, we have in this limit $j_1(x) \simeq x/3$, $y_1(x) \simeq -1/x^2$, $[xj_1(x)]' \simeq 2x/3$ and $[xy_1(x)]' \simeq 1/x^2$ which gives

$$a_1 \simeq \frac{\epsilon_1 \frac{x_1}{3} \frac{2x_0}{3} - \epsilon_0 \frac{x_0}{3} \frac{2x_1}{3}}{\epsilon_1 \frac{x_1}{3} \left( \frac{2x_0}{3} + \frac{i}{x_0^2} \right) - \epsilon_0 \frac{2x_1}{3} \left( \frac{x_0}{3} - \frac{i}{x_0^2} \right)} \simeq \frac{2k_0^3 a^3}{3i} \frac{\epsilon_1 - \epsilon_0}{\epsilon_1 + 2\epsilon_0},$$
(S20a)

$$b_1 \simeq \frac{2k_0^3 a^3}{3i} \frac{\mu_1 - \mu_0}{\mu_1 + 2\mu_0}.$$
(S20b)

We thus get for the polarizabilities

$$\alpha_e \simeq 4\pi \epsilon_0 a^3 \frac{\epsilon_1 - \epsilon_0}{\epsilon_1 + 2\epsilon_0}, \quad \alpha_m \simeq 4\pi a^3 \frac{\mu_1 - \mu_0}{\mu_1 + 2\mu_0},$$
(S21)

which are exactly equal to the results derived using the electro-static and magneto-static approximations method. For a non-magnetic material, $b_1$ becomes

$$b_1 \simeq \frac{x_0^3}{45i} x_0^2 \left( \frac{\epsilon_1}{\epsilon_0} - 1 \right),$$
(S22)

which gives for the magnetic polarizability,

$$\alpha_m \simeq \frac{2\pi}{15} k_0^2 a^5 \left( \frac{\epsilon_1}{\epsilon_0} - 1 \right) = \frac{8\pi^3}{15} a^3 \left( \frac{a}{\lambda} \right)^2 \left( \frac{\epsilon_1}{\epsilon_0} - 1 \right).$$
(S23)

## S2 Vacuum Frictional Torque

### S2.1 Other components of torque

In this section, we provide further discussions of components of the torque other than the $z$ component exerted on a spinning nanosphere near YIG slabs under different bias fields. The $x$ component of torque,

$$
\begin{aligned}
M_x = \frac{\hbar}{4\pi} \int_{-\infty}^{\infty} d\omega \Bigg\{ & \left[2n_1(\omega^-) + 1\right] \left[\text{Im}\left\{\alpha_{m,\perp}(\omega^-)\right\} - \text{Re}\left\{\alpha_{m,g}(\omega^-)\right\}\right] \left[2\text{Im}\left\{G_{zx}^{\text{H}}(\omega)\right\} + 2\text{Re}\left\{G_{zy}^{\text{H}}(\omega)\right\}\right] \\
& -4\left[n_1(\omega) + 1\right] \text{Im}\left\{\alpha_{m,\|}(\omega)\right\} \text{Re}\left\{G_{yz}^{\text{H}}(\omega)\right\} \\
+ \left[2n_0(\omega) + 1\right] \Bigg\{ & \left[\text{Re}\{\alpha_{m,\perp}(\omega^-)\} + \text{Im}\{\alpha_{m,g}(\omega^-)\}\right] \left(\text{Re}\{G_{xz}^{\text{H}}(\omega)\} - \text{Re}\{G_{zx}^{\text{H}}(\omega)\} + \text{Im}\{G_{yz}^{\text{H}}(\omega)\} + \text{Im}\{G_{zy}^{\text{H}}(\omega)\}\right) \\
& + \left[\text{Im}\{\alpha_{m,\perp}(\omega^-)\} - \text{Re}\{\alpha_{m,g}(\omega^-)\}\right] \left(-\text{Im}\{G_{xz}^{\text{H}}(\omega)\} - \text{Im}\{G_{zx}^{\text{H}}(\omega)\} + \text{Re}\{G_{yz}^{\text{H}}(\omega)\} - \text{Re}\{G_{zy}^{\text{H}}(\omega)\}\right) \Bigg\} \\
+ \left[n_0(\omega) + 1\right] \Bigg\{ & -2\text{Re}\{\alpha_{m,\|}(\omega)\} \left(\text{Im}\{G_{zy}^{\text{H}}(\omega)\} + \text{Im}\{G_{yz}^{\text{H}}(\omega)\}\right) + 2\text{Im}\{\alpha_{m,\|}(\omega)\} \left(-\text{Re}\{G_{zy}^{\text{H}}(\omega)\} + \text{Re}\{G_{yz}^{\text{H}}(\omega)\}\right) \Bigg\} \Bigg\},
\end{aligned}
\tag{S24}
$$

and for the $y$ component,

$$
\begin{aligned}
M_y = \frac{\hbar}{4\pi} \int_{-\infty}^{\infty} d\omega \Bigg\{ & \left[2n_1(\omega^-) + 1\right] \left[\text{Im}\left\{\alpha_{m,\perp}(\omega^-)\right\} - \text{Re}\left\{\alpha_{m,g}(\omega^-)\right\}\right] \left[-2\text{Re}\{G_{zx}^{\text{H}}(\omega)\} + 2\text{Im}\{G_{zy}^{\text{H}}(\omega)\}\right] \\
& + 4\left[n_1(\omega) + 1\right] \text{Im}\left\{\alpha_{m,\|}(\omega)\right\} \text{Re}\{G_{xz}^{\text{H}}(\omega)\} \Bigg\} \\
- \left[2n_0(\omega) + 1\right] \Bigg\{ & \left[\text{Re}\{\alpha_{m,\perp}(\omega^-)\} + \text{Im}\{\alpha_{m,g}(\omega^-)\}\right] \left(\text{Im}\{G_{xz}^{\text{H}}(\omega)\} + \text{Im}\{G_{zx}^{\text{H}}(\omega)\} - \text{Re}\{G_{yz}^{\text{H}}(\omega)\} + \text{Re}\{G_{zy}^{\text{H}}(\omega)\}\right) \\
& + \left[\text{Im}\{\alpha_{m,\perp}(\omega^-)\} - \text{Re}\{\alpha_{m,g}(\omega^-)\}\right] \left(\text{Re}\{G_{xz}^{\text{H}}(\omega)\} - \text{Re}\{G_{zx}^{\text{H}}(\omega)\} + \text{Im}\{G_{yz}^{\text{H}}(\omega)\} + \text{Im}\{G_{zy}^{\text{H}}(\omega)\}\right) \Bigg\} \\
- \left[n_0(\omega) + 1\right] \Bigg\{ & -2\text{Re}\alpha_{m,\|}(\omega) \left(\text{Im}\{G_{zx}^{\text{H}}(\omega)\} + \text{Im}\{G_{xz}^{\text{H}}(\omega)\}\right) + 2\text{Im}\alpha_{m,\|}(\omega) \left(-\text{Re}\{G_{zx}^{\text{H}}(\omega)\} + \text{Re}\{G_{xz}^{\text{H}}(\omega)\}\right) \Bigg\} \Bigg\}.
\end{aligned}
\tag{S25}
$$

In the case when the interface is in the $x-y$ plane, we have

$$
\begin{aligned}
\text{Re}\left\{G_{xz}^{\text{H}}(\omega)\right\} = (\pi\omega\rho_0/8)\frac{1}{\pi}\int_0^{2\pi} d\phi \Bigg\{ & \int_0^1 \frac{\kappa^2 d\kappa}{p} \left(\text{Im}\left\{r_{ss}e^{2ik_0pd}\right\} p\cos\phi - \text{Im}\left\{r_{ps}e^{2ik_0pd}\right\} \sin\phi\right) \\
& + \int_1^{\infty} \frac{\kappa^2 d\kappa}{|p|} e^{-2k_0|p|d} \left(\text{Im}\left\{r_{ss}\right\} |p|\cos\phi + \text{Re}\left\{r_{ps}\right\} \sin\phi\right) \Bigg\},
\end{aligned}
\tag{S26a}
$$

$$
\begin{aligned}
\text{Re}\left\{G_{zx}^{\text{H}}(\omega)\right\} = (\pi\omega\rho_0/8)\frac{1}{\pi}\int_0^{2\pi} d\phi \Bigg\{ & \int_0^1 \frac{\kappa^2 d\kappa}{p} \left(-\text{Im}\left\{r_{ss}e^{2ik_0pd}\right\} p\cos\phi - \text{Im}\left\{r_{sp}e^{2ik_0pd}\right\} \sin\phi\right) \\
& + \int_1^{\infty} \frac{\kappa^2 d\kappa}{|p|} e^{-2k_0|p|d} \left(-\text{Im}\left\{r_{ss}\right\} |p|\cos\phi + \text{Re}\left\{r_{sp}\right\} \sin\phi\right) \Bigg\},
\end{aligned}
\tag{S26b}
$$

$$
\begin{aligned}
\text{Im}\left\{G_{xz}^{\text{H}}(\omega)\right\} = (\pi\omega\rho_0/8)\frac{1}{\pi}\int_0^{2\pi} d\phi \Bigg\{ & \int_0^1 \frac{\kappa^2 d\kappa}{p} \left(-\text{Re}\left\{r_{ss}e^{2ik_0pd}\right\} p\cos\phi + \text{Re}\left\{r_{ps}e^{2ik_0pd}\right\} \sin\phi\right) \\
& + \int_1^{\infty} \frac{\kappa^2 d\kappa}{|p|} e^{-2k_0|p|d} \left(-\text{Re}\left\{r_{ss}\right\} |p|\cos\phi + \text{Im}\left\{r_{ps}\right\} \sin\phi\right) \Bigg\},
\end{aligned}
\tag{S26c}
$$

$$\text{Im}\left\{G_{zx}^{\text{H}}(\omega)\right\} = (\pi\omega\rho_0/8)\frac{1}{\pi}\int_0^{2\pi}d\phi\Bigg\{\int_0^1\frac{\kappa^2 d\kappa}{p}\left(\text{Re}\left\{r_{ss}e^{2ik_0pd}\right\}p\cos\phi + \text{Re}\left\{r_{sp}e^{2ik_0pd}\right\}\sin\phi\right)$$
$$+ \int_1^\infty\frac{\kappa^2 d\kappa}{|p|}e^{-2k_0|p|d}\left(\text{Re}\left\{r_{ss}\right\}|p|\cos\phi + \text{Im}\left\{r_{sp}\right\}\sin\phi\right)\Bigg\}, \tag{S26d}$$

and

$$\text{Re}\left\{G_{zy}^{\text{H}}(\omega)\right\} = (\pi\omega\rho_0/8)\frac{1}{\pi}\int_0^{2\pi}d\phi\Bigg\{\int_0^1\frac{\kappa^2 d\kappa}{p}\left(-\text{Im}\left\{r_{ss}e^{2ik_0pd}\right\}p\sin\phi + \text{Im}\left\{r_{sp}e^{2ik_0pd}\right\}\cos\phi\right)$$
$$+ \int_1^\infty\frac{\kappa^2 d\kappa}{|p|}e^{-2k_0|p|d}\left(-\text{Im}\left\{r_{ss}\right\}|p|\sin\phi - \text{Re}\left\{r_{sp}\right\}\cos\phi\right)\Bigg\}, \tag{S27a}$$

$$\text{Re}\left\{G_{yz}^{\text{H}}(\omega)\right\} = (\pi\omega\rho_0/8)\frac{1}{\pi}\int_0^{2\pi}d\phi\Bigg\{\int_0^1\frac{\kappa^2 d\kappa}{p}\left(\text{Im}\left\{r_{ss}e^{2ik_0pd}\right\}p\sin\phi + \text{Im}\left\{r_{ps}e^{2ik_0pd}\right\}\cos\phi\right)$$
$$+ \int_1^\infty\frac{\kappa^2 d\kappa}{|p|}e^{-2k_0|p|d}\left(\text{Im}\left\{r_{ss}\right\}|p|\sin\phi - \text{Re}\left\{r_{ps}\right\}\cos\phi\right)\Bigg\}, \tag{S27b}$$

$$\text{Im}\left\{G_{zy}^{\text{H}}(\omega)\right\} = (\pi\omega\rho_0/8)\frac{1}{\pi}\int_0^{2\pi}d\phi\Bigg\{\int_0^1\frac{\kappa^2 d\kappa}{p}\left(\text{Re}\left\{r_{ss}e^{2ik_0pd}\right\}p\sin\phi - \text{Re}\left\{r_{sp}e^{2ik_0pd}\right\}\cos\phi\right)$$
$$+ \int_1^\infty\frac{\kappa^2 d\kappa}{|p|}e^{-2k_0|p|d}\left(\text{Re}\left\{r_{ss}\right\}|p|\sin\phi - \text{Im}\left\{r_{sp}\right\}\cos\phi\right)\Bigg\}, \tag{S27c}$$

$$\text{Im}\left\{G_{yz}^{\text{H}}(\omega)\right\} = (\pi\omega\rho_0/8)\frac{1}{\pi}\int_0^{2\pi}d\phi\Bigg\{\int_0^1\frac{\kappa^2 d\kappa}{p}\left(-\text{Re}\left\{r_{ss}e^{2ik_0pd}\right\}p\sin\phi - \text{Re}\left\{r_{ps}e^{2ik_0pd}\right\}\cos\phi\right)$$
$$+ \int_1^\infty\frac{\kappa^2 d\kappa}{|p|}e^{-2k_0|p|d}\left(-\text{Re}\left\{r_{ss}\right\}|p|\sin\phi - \text{Im}\left\{r_{ps}\right\}\cos\phi\right)\Bigg\}. \tag{S27d}$$

And for the case when it is in the $x-z$ plane

$$\text{Re}\left\{G_{xz}^{\text{H}}(\omega)\right\} = (\pi\omega\rho_0/8)\frac{1}{\pi}\int_0^{2\pi}d\phi\Bigg\{\int_0^1\frac{\kappa d\kappa}{2p}\Big[\sin 2\phi\text{Im}\left\{r_{pp}e^{2ik_0pd}\right\} + p^2\sin 2\phi\text{Im}\left\{r_{ss}e^{2ik_0pd}\right\}$$
$$+2p\text{Im}\left\{r_{ps}e^{2ik_0pd}\right\}\cos^2\phi + 2p\text{Im}\left\{r_{sp}e^{2ik_0pd}\right\}\sin^2\phi\Big]$$
$$+\int_1^\infty\frac{\kappa d\kappa}{2|p|}\Big[-\sin 2\phi\left(1 + \text{Re}\left\{r_{pp}\right\}e^{-2k_0|p|d}\right) - p^2\sin 2\phi\text{Re}\left\{r_{ss}\right\}e^{-2k_0|p|d}$$
$$+2\text{Im}\{r_{ps}\}|p|e^{-2k_0|p|d}\cos^2\phi + 2\text{Im}\{r_{sp}e^{-2k_0|p|d}\}\sin^2\phi\Big]\Bigg\}, \tag{S28a}$$

$$\text{Re}\left\{G_{zx}^{\text{H}}(\omega)\right\} = (\pi\omega\rho_0/8)\frac{1}{\pi}\int_0^{2\pi}d\phi\Bigg\{\int_0^1\frac{\kappa d\kappa}{2p}\Big[\sin 2\phi\text{Im}\left\{r_{pp}e^{2ik_0pd}\right\} + p^2\sin 2\phi\text{Im}\left\{r_{ss}e^{2ik_0pd}\right\}$$
$$-2p\text{Im}\left\{r_{ps}e^{2ik_0pd}\right\}\sin^2\phi - 2p\text{Im}\left\{r_{sp}e^{2ik_0pd}\right\}\cos^2\phi\Big]$$
$$+\int_1^\infty\frac{\kappa d\kappa}{2|p|}\Big[-\sin 2\phi\left(1 + \text{Re}\left\{r_{pp}\right\}e^{-2k_0|p|d}\right) - p^2\sin 2\phi\text{Re}\left\{r_{ss}\right\}e^{-2k_0|p|d}$$
$$-2\text{Im}\{r_{ps}\}|p|e^{-2k_0|p|d}\sin^2\phi - 2\text{Im}\{r_{sp}\}|p|e^{-2k_0|p|d}\cos^2\phi\Big]\Bigg\}, \tag{S28b}$$

$$
\begin{aligned}
\mathrm{Im}\left\{G_{xz}^{\mathrm{H}}(\omega)\right\} = (\pi\omega\rho_0/8)\frac{1}{\pi}\int_0^{2\pi}d\phi\Bigg\{ &\int_0^1\frac{\kappa d\kappa}{2p}\Big[-\sin 2\phi\left(1+\mathrm{Re}\left\{r_{pp}e^{2ik_0 pd}\right\}\right)-p^2\sin 2\phi\,\mathrm{Re}\left\{r_{ss}e^{2ik_0 pd}\right\} \\
&-2p\,\mathrm{Re}\left\{r_{ps}e^{2ik_0 pd}\right\}\cos^2\phi-2p\,\mathrm{Re}\left\{r_{sp}e^{2ik_0 pd}\right\}\sin^2\phi\Big] \\
&+\int_1^{\infty}\frac{\kappa d\kappa}{2|p|}e^{-2k_0|p|d}\Big[-\sin 2\phi\,\mathrm{Im}\left\{r_{pp}\right\}-p^2\sin 2\phi\,\mathrm{Im}\{r_{ss}\} \\
&-2\mathrm{Re}\{r_{ps}\}|p|\cos^2\phi-2\mathrm{Re}\{r_{sp}\}|p|\sin^2\phi\Big]\Bigg\},
\end{aligned}
$$

(S28c)

$$
\begin{aligned}
\mathrm{Im}\left\{G_{zx}^{\mathrm{H}}(\omega)\right\} = (\pi\omega\rho_0/8)\frac{1}{\pi}\int_0^{2\pi}d\phi\Bigg\{ &\int_0^1\frac{\kappa d\kappa}{2p}\Big[-\sin 2\phi\left(1+\mathrm{Re}\left\{r_{pp}e^{2ik_0 pd}\right\}\right)-p^2\sin 2\phi\,\mathrm{Re}\left\{r_{ss}e^{2ik_0 pd}\right\} \\
&+2p\,\mathrm{Re}\left\{r_{ps}e^{2ik_0 pd}\right\}\sin^2\phi+2p\,\mathrm{Re}\left\{r_{sp}e^{2ik_0 pd}\right\}\cos^2\phi\Big] \\
&+\int_1^{\infty}\frac{\kappa d\kappa}{2|p|}e^{-2k_0|p|d}\Big[-\sin 2\phi\,\mathrm{Im}\left\{r_{pp}\right\}-p^2\sin 2\phi\,\mathrm{Im}\{r_{ss}\} \\
&+2\mathrm{Re}\{r_{ps}\}|p|\sin^2\phi+2\mathrm{Re}\{r_{sp}\}|p|\cos^2\phi\Big]\Bigg\},
\end{aligned}
$$

(S28d)

and the expressions for the real and imaginary parts of $G_{zy}^{\mathrm{H}}$ and $G_{yz}^{\mathrm{H}}$ are the same as the ones for $G_{xz}^{\mathrm{H}}$ and $G_{zx}^{\mathrm{H}}$, respectively, for when the interface is in the $x-y$ plane as given in Eq. (S26). We can find the $x$ and $y$ components of torque by plugging these expressions into Eqs. (S24) and (S25) for the two cases when the interface is the $x-y$ or $x-z$ plane. We present the plots of these torques at the end of this section.

### S2.2 Recoil torque

There is also another contribution to the torque from the case when the induced dipole moments on the YIG sphere re-radiate due to the vacuum electric field fluctuations. This causes a recoil torque on the sphere and can be written as

$$
\boldsymbol{M}^{\mathrm{rec}} = \langle\boldsymbol{m}^{\mathrm{ind}}\times\boldsymbol{H}^{\mathrm{sc}}\rangle,
$$

(S29)

where $\boldsymbol{H}^{\mathrm{sc}}$ is the scattered fields from the dipole and are given by,

$$
\boldsymbol{H}^{\mathrm{sc}}(\boldsymbol{r},\omega) = \bar{\boldsymbol{G}}^{\mathrm{H}}(\boldsymbol{r},\boldsymbol{r}',\omega)\cdot\boldsymbol{m}^{\mathrm{ind}}(\boldsymbol{r}',\omega),
$$

(S30)

which shows that this term is of higher order contribution and is thus smaller than the torque discussed in the main text. Repeating a similar procedure used before and plugging in all of the induced terms and writing them in terms of the fluctuations, we find after some algebra,

$$
\begin{aligned}
M_z^{\mathrm{rec}} =& \frac{\hbar}{\pi}\int_{-\infty}^{\infty}d\omega[n_0(\omega)+1] \\
&\Bigg\{\mathrm{Im}\{G_{xx}\}\left[\mathrm{Re}\{G_{yx}\}\alpha_{\perp\perp}^{\mathrm{eff}}-\mathrm{Re}\{G_{xy}\}\alpha_{gg}^{\mathrm{eff}}+\mathrm{Re}\{\alpha_{\perp g}\}\mathrm{Re}\{G_{yy}-G_{xx}\}+\mathrm{Im}\{\alpha_{\perp g}\}\mathrm{Im}\{G_{yy}+G_{xx}\}\right] \\
&+\mathrm{Im}\{G_{yy}\}\left[\mathrm{Re}\{G_{yx}\}\alpha_{gg}^{\mathrm{eff}}-\mathrm{Re}\{G_{xy}\}\alpha_{\perp\perp}^{\mathrm{eff}}+\mathrm{Re}\{\alpha_{\perp g}\}\mathrm{Re}\{G_{xx}-G_{yy}\}+\mathrm{Im}\{\alpha_{\perp g}\}\mathrm{Im}\{G_{yy}+G_{xx}\}\right] \\
&+\mathrm{Re}\{G_{yx}-G_{xy}\}\left[\mathrm{Re}\{G_{yx}-G_{xy}\}\mathrm{Im}\{\alpha_{\perp g}\}+\frac{1}{2}\mathrm{Im}\{G_{yy}+G_{xx}\}\left(\alpha_{\perp\perp}+\alpha_{gg}\right)\right] \\
&+\mathrm{Im}\{G_{yx}+G_{xy}\}\left[-\mathrm{Re}\{G_{yx}+G_{xy}\}\mathrm{Re}\{\alpha_{\perp g}\}+\frac{1}{2}\mathrm{Re}\{G_{xx}-G_{yy}\}\left(\alpha_{gg}-\alpha_{\perp\perp}\right)\right] \\
&+\frac{1}{2}\mathrm{Im}\Big\{\alpha_{m,\parallel}^{\mathrm{eff}*}\left[\left(G_{xz}-G_{zx}^*\right)\left(G_{yz}^*\alpha_{m,\perp}^{\mathrm{eff}}-G_{zx}^*\alpha_{m,g}^{\mathrm{eff}}\right)-\left(G_{yz}-G_{zy}^*\right)\left(G_{yz}^*\alpha_{m,g}^{\mathrm{eff}}+G_{zx}^*\alpha_{m,\perp}^{\mathrm{eff}}\right)\right]\Big\}\Bigg\},
\end{aligned}
$$

(S31)

where we have defined

$$
\begin{aligned}
\alpha_{m,\perp\perp}^{\text{eff}}(\omega) &= \alpha_{m,\perp}^{\text{eff}}(\omega)\alpha_{m,\perp}^{\text{eff}}(-\omega), \quad &\alpha_{m,gg}^{\text{eff}}(\omega) &= \alpha_{m,g}^{\text{eff}}(\omega)\alpha_{m,g}^{\text{eff}}(-\omega), \\
\alpha_{m,\perp g}^{\text{eff}}(\omega) &= \alpha_{m,\perp}^{\text{eff}}(\omega)\alpha_{m,g}^{\text{eff}}(-\omega), \quad &\alpha_{m,g\perp}^{\text{eff}}(\omega) &= \alpha_{m,\perp}^{\text{eff}}(-\omega)\alpha_{m,g}^{\text{eff}}(\omega),
\end{aligned}
\tag{S32}
$$

and have used the facts that $\alpha_{m,\perp\perp}^{\text{eff}}(\omega)$ and $\alpha_{m,gg}^{\text{eff}}(\omega)$ are real, and $\alpha_{m,\perp g}^{\text{eff}}(\omega) = \left[\alpha_{m,g\perp}^{\text{eff}}(\omega)\right]^*$.

Note that we have dropped the frequency dependence as well as the H superscript of the Green's function in Eq. (S31) for simplicity. For the special case when the substrate material is isotropic, the non-diagonal elements of the Green's function become zero, and we get

$$
M_z^{\text{rec}} = \frac{\hbar}{\pi}\int_{-\infty}^{\infty} d\omega\left[n_0(\omega)+1\right]\left\{\text{Im}\{G_{xx}-G_{yy}\}\text{Re}\{G_{yy}-G_{xx}\}\text{Re}\{\alpha_{\perp g}\}+\left[\text{Im}\{G_{xx}+G_{yy}\}\right]^2\text{Im}\{\alpha_{\perp g}\}\right\}.
\tag{S33}
$$

Note that the expressions for the real and imaginary parts of $G_{xz}$ and $G_{yz}$ are given by Eqs. (S26),(S27), and (S28) for the two possible interface directions while the imaginary parts of $G_{xx}$ and $G_{yy}$ are given by equations in Appendix B. Also note that $\text{Re}\left\{G_{yz}^{\text{H}}\right\}$ for when the interface is the $x-y$ plane is the same as $\text{Re}\left\{G_{xz}^{\text{H}}\right\}$ for when the interface is in the $x-z$ plane given by Eq. (S28). Also $\text{Re}\left\{G_{yx}^{\text{H}}\right\}$ for when the interface is the $x-z$ plane is the same as $\text{Re}\left\{G_{zy}^{\text{H}}\right\}$ for when the interface is in the $x-y$ plane given by Eq. (S27). Thus, the only new term is $\text{Re}\{G_{yy}-G_{xx}\}$ which is given by

$$
\begin{aligned}
\text{Re}\left\{G_{yy}^{\text{H}}(\omega)-G_{xx}^{\text{H}}(\omega)\right\} =& (\pi\omega\rho_0/8)\frac{1}{\pi}\int_0^{2\pi}d\phi\left\{\int_0^1\frac{\kappa d\kappa}{p}\left[-\cos 2\phi\text{Im}\left\{r_{pp}e^{2ik_0pd}\right\}-\cos 2\phi\text{Im}\left\{r_{ss}e^{2ik_0pd}\right\}\right.\right. \\
&\left.+2p\sin\phi\cos\phi\text{Im}\left\{(r_{ps}-r_{sp})e^{2ik_0pd}\right\}\right] \\
&+\int_1^{\infty}\frac{\kappa d\kappa}{|p|}\left[\cos 2\phi\left(\kappa^2+\text{Re}\{r_{pp}\}e^{-2k_0|p|d}\right)+p^2\cos 2\phi\text{Re}\{r_{ss}\}e^{-2k_0|p|d}\right. \\
&\left.\left.+2|p|\sin\phi\cos\phi\text{Im}\{r_{ps}-r_{sp}\}e^{-2k_0|p|d}\right]\right\},
\end{aligned}
\tag{S34}
$$

when the interface is the $x-y$ plane, and

$$
\begin{aligned}
\text{Re}\left\{G_{yy}^{\text{H}}(\omega)-G_{xx}^{\text{H}}(\omega)\right\} =& (\pi\omega\rho_0/8)\frac{1}{\pi}\int_0^{2\pi}d\phi\left\{\int_0^1\frac{\kappa d\kappa}{p}\left[\cos^2\phi\text{Im}\left\{r_{pp}e^{2ik_0pd}\right\}-\left(\kappa^2+p^2\sin^2\phi\right)\text{Im}\left\{r_{ss}e^{2ik_0pd}\right\}\right.\right. \\
&\left.-p\sin\phi\cos\phi\text{Im}\left\{(r_{ps}-r_{sp})e^{2ik_0pd}\right\}\right] \\
&+\int_1^{\infty}\frac{\kappa d\kappa}{|p|}\left[-\cos^2\phi\left(1+\text{Re}\{r_{pp}\}e^{-2k_0|p|d}\right)-p^2\sin^2\phi+\kappa^2+\left(\kappa^2+p^2\sin^2\phi\right)\text{Re}\{r_{ss}\}e^{-2k_0|p|d}\right. \\
&\left.\left.-|p|\sin\phi\cos\phi\text{Im}\{r_{ps}-r_{sp}\}e^{-2k_0|p|d}\right],\right.
\end{aligned}
\tag{S35}
$$

when the interface is the $x-z$ plane.

### S2.3 Plots of torque terms

In this section, we present the components of torque derived in previous sections for YIG slabs with various bias magnetic fields and for the two cases when the slab is the $x-y$ and $x-z$ planes.

Figure S2 shows the plots of $M_x$, $M_y$, $M_z$, and $M_{\text{rec}}$ derived in the previous sections for the magnetic and electric fluctuations. The expressions for the torques due to the electric fields and dipoles fluctuations are found by changing $s$ to $p$ and $p$ to $s$ in $r_{ss}$, $r_{pp}$, $r_{sp}$, and $r_{ps}$, in the expressions for the Green's functions. Moreover, magnetic polarizability is replaced by a simple isotropic electric polarizability, assuming a simple dielectric polarizability scalar for the YIG and Al interfaces. The results are for three directions of the bias magnetic field for the YIG interface labeled as $x-$, $y-$, and $z-$bias. The meaning of these bias directions is demonstrated in Fig. S1 when the YIG slab is the $x-y$ and $x-z$ planes.

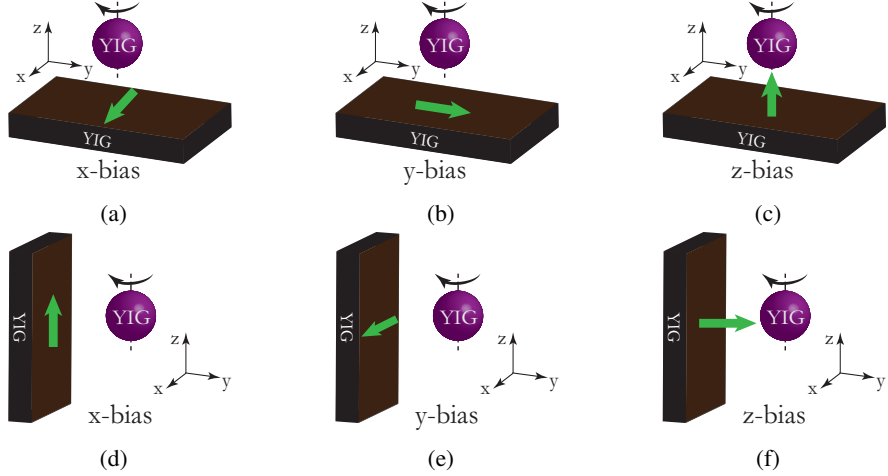

Figure S1: Schematics of different bias directions for the YIG interface for the two cases of the interface being the $x - y$ (top row) and $x - z$ planes (bottom row). The green arrow shows the direction of the bias magnetic field applied to the slab of YIG.

It is interesting to note that in Figs. (S2a), (S2e), and (S2g), the sphere can experience a large value of torque along $x$ or $y$ directions for the $x-$ or $y-$biases. This means that in these cases, the sphere can rotate out of the rotation axis and start to precess. This will, of course, change the validity of the equations found for the vacuum radiation and frictional torque along the $z$ axis since it has been assumed that the sphere is always rotating around the $z$ axis and is also magnetized along that axis. This torque is still small enough compared to the driving torque of the trapping laser and it will still give enough time to make the observations. A more careful investigation of these components of torque is out of the scope of this study and will be explored in the future.

Figures S2i-S2p show the axial torque $M_z$ as well as the recoil torque $M_{rec}$ for all orientations of the bias magnetic field and YIG slab. As expected, the recoil torque is much smaller than $M_z$ since it is a second-order term.

Figure S3 shows the results for $M_z$ and $M_{rec}$ for the case when the Al interface is placed in the vicinity of the spinning sphere. Because Al is an isotropic material, $M_x$ and $M_y$ vanish for both orientations of the interface and thus are not included in the plots of the torques. Note that similar to the YIG interface results, $M_{rec}$ is much smaller than the $M_z$ for all cases of the Al interface. These results show that the recoil torque $M_{rec}$ can be ignored in all studied cases.

## S3    Experimental Considerations

In this section, we present details of the experimental analysis regarding negative torque and shot noise heating due to the trapping laser discussed in Appendix H.

### S3.1    Effect of torque due to the trapping laser

When the trapping laser is linearly polarized, it can exert a negative torque on the spinning particle. The torque on the sphere due to the laser is given by $\boldsymbol{M}_{opt} = \frac{1}{2}\mathrm{Re}\{\boldsymbol{p}^* \times \boldsymbol{E}\}$ [1], where $\boldsymbol{p}$ is the dipole moment of the sphere, given by $\boldsymbol{p} = \bar{\boldsymbol{\alpha}}^{eff} \cdot \boldsymbol{E}$, with $\bar{\boldsymbol{\alpha}}^{eff}$ being the effective polarizability of the sphere as seen in the frame of the lab, and $\boldsymbol{E}$ is the electric field from the laser. As shown in Appendix A, the polarizability tensor of the sphere when it is spinning in the $x - y$ plane is given by

$$\bar{\boldsymbol{\alpha}}^{eff}(\omega) = \begin{pmatrix} \alpha_{\perp}^{eff}(\omega) & -\alpha_{g}^{eff}(\omega) & 0 \\ \alpha_{g}^{eff}(\omega) & \alpha_{\perp}^{eff}(\omega) & 0 \\ 0 & 0 & \alpha_{\parallel}^{eff}(\omega) \end{pmatrix}, \tag{S36}$$

where

$$\alpha_{\perp}^{eff}(\omega) = \frac{1}{2}\left[\alpha(\omega^+) + \alpha(\omega^-)\right], \quad \alpha_g(\omega) = -\frac{i}{2}\left[\alpha(\omega^+) - \alpha(\omega^-)\right], \quad \alpha_{\parallel}(\omega) = \alpha(\omega), \tag{S37}$$

with $\alpha(\omega)$ being the electric polarizability of YIG at the laser frequency. Note that here, we have assumed that the polarizability of the YIG is scalar in the range of frequencies around 1550 nm. Plugging these into the equation for

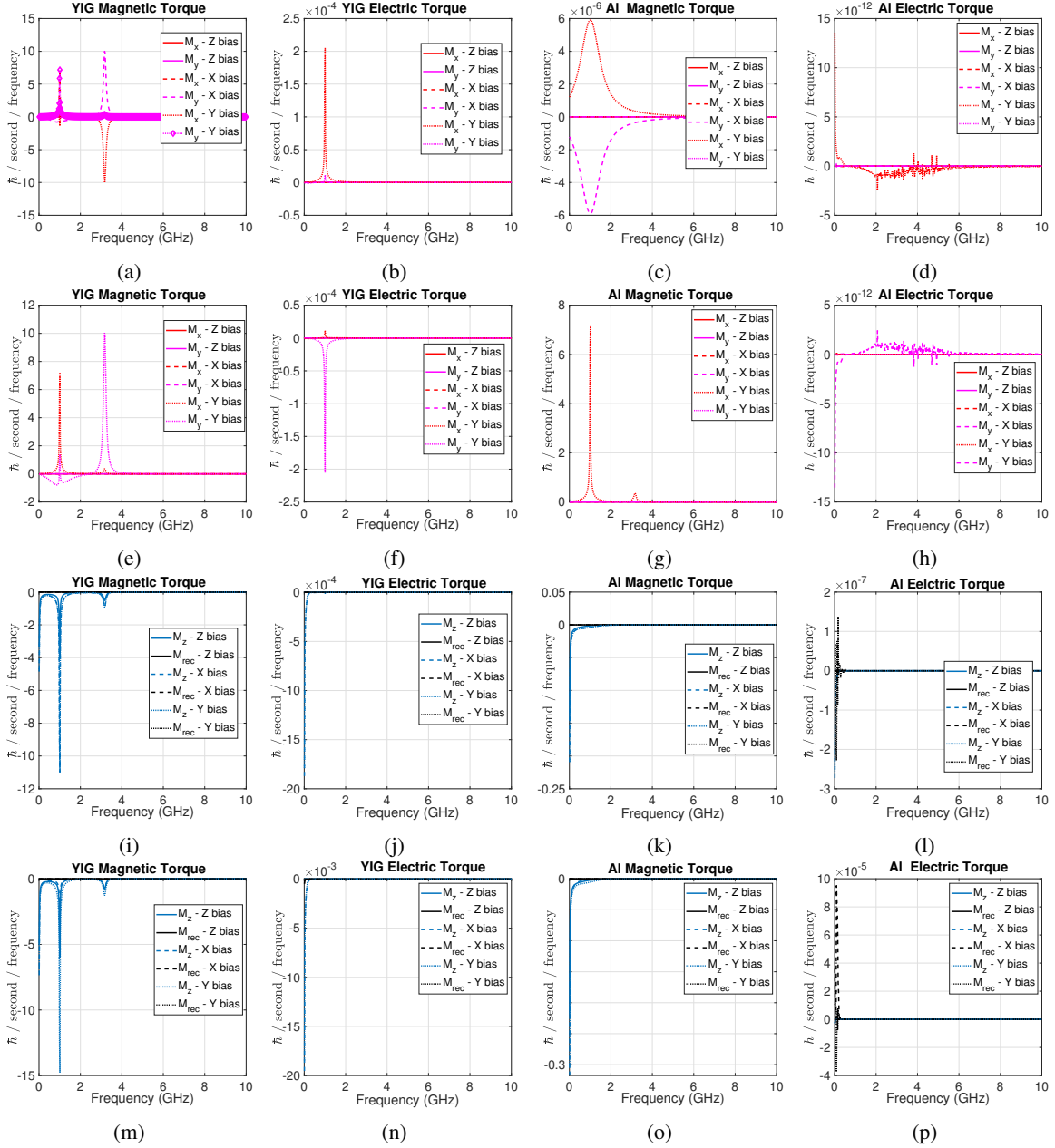

Figure S2: Plots of $M_x$ and $M_y$ (first two rows) and $M_z$ and $M_{\text{rec}}$ (second two rows) in the vicinity of the YIG slab when the slab is the $x-y$ plane (first and third rows), and when the slab is $x-z$ plane (second and fourth rows). The plots show the results for various magnetic field directions. The meanings of $x-$, $y-$, and $z-$bias are demonstrated in Fig. S1 for the two orientations of the interface.

the exerted torque, we find the $z$ component of the torque

$$
\begin{aligned}
M_{\text{opt}} &= \frac{1}{2} \text{Re} \left\{ \alpha_{\perp}^{\text{eff}*}(\omega) E_x^* E_y - \alpha_g^{\text{eff}*}(\omega) E_y^* E_y - \alpha_g^{\text{eff}*}(\omega) E_x^* E_x - \alpha_{\perp}^{\text{eff}*}(\omega) E_y^* E_x \right\} \\
&= \frac{1}{2} \left[ \text{Im}\{\alpha_{\perp}^{\text{eff}}(\omega)\} \text{Im}\{\boldsymbol{E}^* \times \boldsymbol{E}\} - \text{Re}\{\alpha_g^{\text{eff}}(\omega)\} \left( |E_x|^2 + |E_y|^2 \right) \right] \\
&= \frac{1}{2} \left[ \text{Im}\{\alpha(\omega^+) + \alpha(\omega^-)\} \text{Im}\{\boldsymbol{E}^* \times \boldsymbol{E}\} - \text{Im}\{\alpha(\omega^+) - \alpha(\omega^-)\} \left( |E_x|^2 + |E_y|^2 \right) \right].
\end{aligned} \tag{S38}
$$

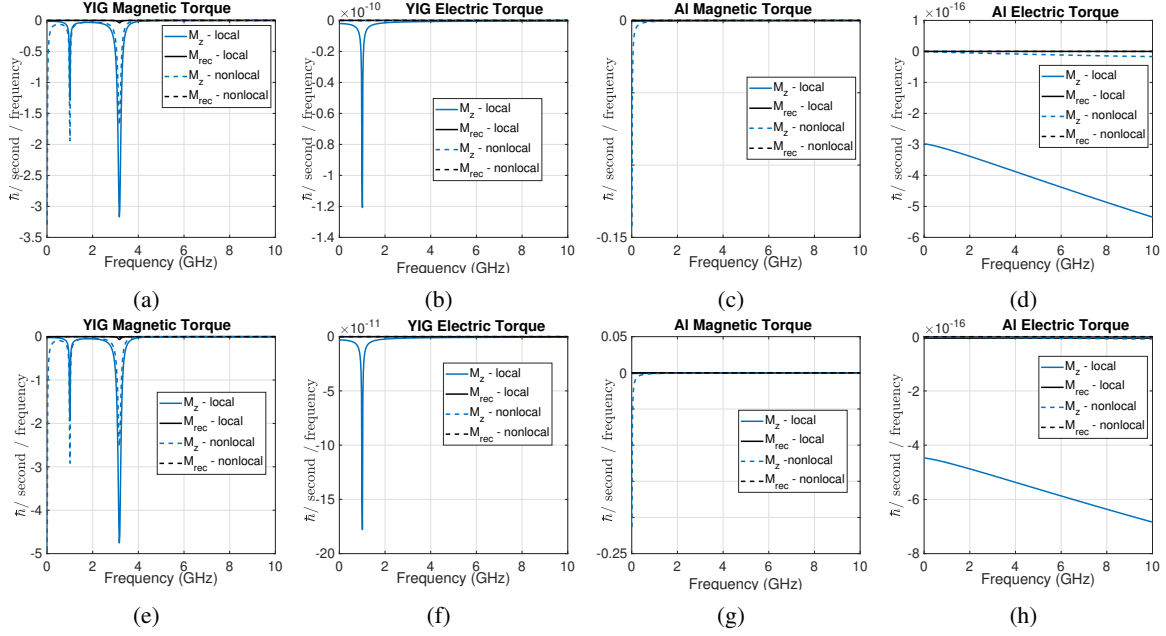

Figure S3: Plots of $M_z$ and $M_{\rm rec}$ in the vicinity of the YIG slab when the slab is the $x-y$ plane (first row) and when the slab is $x-z$ plane (second row). Note that due to the isotropy of Al, the other components of torque, including $M_x$ and $M_y$, vanish.

The first term is proportional to the spin of the electromagnetic field and causes a positive torque on the particle. This is the term for the transferring of angular momentum from the laser to the particle. The second term is negative and thus causes a negative torque on the sphere. In the case when the laser is linearly polarized, this negative term is proportional to ${\rm Im}\{\alpha(\omega_0 + \Omega)\} - {\rm Im}\{\alpha(\omega_0 - \Omega)\}$ where $\omega_0 = 1.21 \times 10^{16}$ is the frequency of the laser, and $\Omega = 6.28 \times 10^9$ is the rotation frequency. Since $\Omega \ll \omega_0$, we get $\alpha(\omega^+) \simeq \alpha(\omega^-)$ and thus the second term is negligible. We can thus ignore the negative torque coming from the laser when the laser is linearly polarized.

### S3.2 Effect of heating due to the shot noise

The particle can heat up due to the shot noise of the trapping laser [4]. In this section, we calculate the rate of temperature change due to the shot noise and vacuum radiation, respectively. The rate of energy change in the nanosphere due to the shot noise is [4],

$$\dot{E}_{T_R} = \frac{\hbar k}{M} \frac{I_L}{c} \sigma, \tag{S39}$$

where $\omega_L = ck$ is the laser frequency, $I_L$ is the power of the laser per unit area, $M$ is the mass of the particle, and $\sigma$ is the cross section of scattering where, which is equal to $\sigma = \left(\frac{8\pi}{3}\right)\left(\frac{\alpha k^2}{4\pi\epsilon}\right)^2$ for Rayleigh particles with the polarizability $\alpha = 4\pi\epsilon_0 a^3\left(\frac{\epsilon-1}{\epsilon+2}\right)$. For the range of wavelengths around visible and infrared, the Rayleigh limit is valid for particles of radii $a$ smaller than 50nm. Since the radius of the particle in our case is 200 nm, this expression may not be valid. Mie scattering parameters should be used to evaluate the scattering cross section. Assuming the trapping laser wavelength of $\lambda = 1550$ nm and using the Mie theory, the rate of energy change of YIG with refractive index $n = 2.21$ [3] is close to that of the diamond with n = 2.39 in the Rayleigh limit [4]. Therefore, we get the energy change rate in the sphere

$$\dot{E}_{T_R} = \frac{2\hbar\omega_0}{\rho c^2 A} P_0 a^3 k^4 \left(\frac{n^2-1}{n^2+2}\right)^2, \tag{S40}$$

where $A = \pi R_L^2$ is the area of the beam where the laser with the power $P_0$ is focused on, and $\rho$ is the mass density which for YIG is $\rho = 5110{\rm kg/m^3}$. For a laser power of 500 mW focused on an area of radius $0.7566\mu{\rm m}$, we find

$$\dot{T}_L = 15.45 K/s. \tag{S41}$$

This is a very small temperature change compared to the time scale of the rotation, which is 1 ns. Therefore, the thermodynamic equilibrium condition for the FDT is still valid. This temperature change gets damped by the radiated

power of the sphere due to the rotation. For a YIG sphere spinning at about $0.5\mu$m from the aluminum interface, the rate of change due to vacuum radiation at the equilibrium temperature $T_0 = 300$ K is,

$$\dot{T}_R = -362.973 K/s, \tag{S42}$$

which is much larger than the temperature rise due to the shot noise of the laser, and this shows that the sphere will cool down. Note that this energy heats the aluminum instead. In this derivation, we have not included the heating due to the noise in the aluminum or YIG interface. The value found in Eq. (S42) is much smaller at lower temperatures.



\end{document}